\documentclass[twocolumn,a4paper,reprint,amssymb,aps,prd,groupedaddress,superscriptaddress,preprintnumbers,amsmath,nobibnotes,showpacs,nofootinbib]{revtex4-2}
\usepackage{booktabs} 
\usepackage{lipsum} 
\usepackage{graphicx}
\usepackage{color}
\usepackage{amsfonts,latexsym}
\usepackage{bm}
\usepackage{cancel}
\usepackage{epsfig}
\usepackage{ulem}
\usepackage{comment}
\usepackage{lipsum}  

\usepackage{amsmath}
\usepackage{hhline} 
\usepackage{enumerate}
\usepackage[utf8]{inputenc}
\usepackage[dvipsnames]{xcolor}
\usepackage[breaklinks=true,colorlinks=true,linkcolor=blue,urlcolor=Blue,citecolor=MidnightBlue,
bookmarks=true,bookmarksopenlevel=2]{hyperref}
\usepackage{bm}
\usepackage{amsmath}
\usepackage{amssymb}
\usepackage{academicons}
\usepackage{verbatim}
\usepackage[capitalize]{cleveref}
\usepackage{soul}
\usepackage{multirow}
\usepackage{enumitem}

\usepackage{array}
\newcommand{\PreserveBackslash}[1]{\let\temp=\\#1\let\\=\temp}
\newcolumntype{C}[1]{>{\PreserveBackslash\centering}p{#1}}
\newcolumntype{R}[1]{>{\PreserveBackslash\raggedleft}p{#1}}
\newcolumntype{L}[1]{>{\PreserveBackslash\raggedright}p{#1}}

\newcommand{\be}{\begin{equation}}
\newcommand{\ee}{\end{equation}}
\newcommand{\ba}{\begin{eqnarray}}
\newcommand{\ea}{\end{eqnarray}}

\newcommand{\bea}{\begin{eqnarray*}}
\newcommand{\eea}{\end{eqnarray*}}

\newcommand{\orcid}[1]{\href{https://orcid.org/#1}{\textcolor[HTML]{A6CE39}{\aiOrcid}}}

\begin{document}

\title{Dynamical dark energy with AdS-to-dS and dS-to-dS transitions:\\
Implications for the $H_0$ tension}

\author{\"{O}zg\"{u}r Akarsu}
\email{akarsuo@itu.edu.tr} 
\affiliation{Department of Physics, Istanbul Technical University, Maslak 34469 Istanbul, T\"{u}rkiye}

\author{Leandros Perivolaropoulos}
\email{leandros@uoi.gr}
\affiliation{Department of Physics, University of Ioannina, GR-45110, Ioannina, Greece}

\author{Anna Tsikoundoura}
\email{antsik7@gmail.com }
\affiliation{Department of Physics, University of Ioannina, GR-45110, Ioannina, Greece}

\author{A. Emrah Y\"{u}kselci} 
\email{yukselcia@itu.edu.tr}
\affiliation{Department of Physics, Istanbul Technical University, Maslak 34469 Istanbul, T\"{u}rkiye}

\author{Alexander Zhuk}
\email{ai.zhuk2@gmail.com}
\affiliation{Center for Advanced Systems Understanding, Untermarkt 20, 02826 G\"{o}rlitz, Germany}
\affiliation{Helmholtz-Zentrum Dresden-Rossendorf, Bautzner Landstra\ss e 400, 01328 Dresden, Germany}
\affiliation{Astronomical Observatory, Odesa I.I. Mechnikov National University, Dvoryanskaya St. 2, Odesa 65082, Ukraine} 


\begin{abstract}
We investigate the dynamics and cosmological implications of dynamical dark energy (DE), modeled as a scalar field with a hyperbolic tangent potential that induces a smooth shift in the effective cosmological constant (CC), encompassing transitions such as AdS-to-dS, 0-to-dS, and dS-to-dS, with the mirror AdS-to-dS as a particular case aligned with the $\Lambda_{\rm s}$CDM scenario. In our construction, a phantom scalar field with a negative kinetic term drives a bottom-up transition from an anti–de Sitter (AdS)–like vacuum at high redshifts to a de Sitter (dS)–like vacuum at low redshifts, thereby providing a physical underpinning for the phenomenologically successful $\Lambda_{\rm s}$CDM scenario. Despite the negative kinetic term, the step-like form of the potential prevents pathologies such as unbounded energy growth, Big Rip singularities, and violations of the weak energy condition (WEC). Our numerical integration of the equations of motion shows that the model is consistent with both CMB data and the local SH0ES determination of $H_0$, thereby addressing the $H_0$ tension, with all key kinematical parameters—$H(z)$, $\dot{H}(z)$, and $q(z)$—evolving smoothly and continuously. The total energy density of the combined phantom and matter system remains positive at all times, and the effective equation of state (EoS) stays above $-1$, ensuring that the WEC is satisfied. While the phantom field’s energy density and pressure remain finite throughout, its EoS exhibits a safe singularity as its energy density smoothly crosses zero. We perform a detailed analysis of the transition period, demonstrating that the evolution of the DE density from a negative CC–like regime to a positive one does not exactly mirror the behavior of the potential---for instance, it lasts longer---as it also involves the kinetic term. We also show that analogous quintessence models featuring dS-to-dS transitions predict an $H_0$ value lower than that of Planck-$\Lambda$CDM, thereby failing to address the $H_0$ tension. Our results establish a robust theoretical foundation for $\Lambda_{\rm s}$CDM and underscore its potential as a physically motivated framework for addressing major cosmological tensions.

\end{abstract}

\maketitle

\section{Introduction}
\label{sec:Introduction}

The Hubble constant ($H_0$) tension has emerged as the most significant challenge in modern precision cosmology~\cite{Verde:2019ivm,DiValentino:2020zio,DiValentino:2021izs,Perivolaropoulos:2021jda,Schoneberg:2021qvd,Shah:2021onj,Abdalla:2022yfr,DiValentino:2022fjm,Kamionkowski:2022pkx,Hu:2023jqc,Verde:2023lmm,DiValentino:2024yew,Perivolaropoulos:2024yxv}. While the longstanding cosmological constant (CC) problem~\cite{Weinberg:1988cp,Peebles:2002gy} has motivated decades of theoretical inquiry, the $H_0$ tension is particularly striking because it pertains to low-energy physics—a domain once assumed to be well understood. This tension may reveal new physics beyond the standard $\Lambda$CDM framework—assuming it is not merely an artifact of unidentified systematics—and offer profound insights into both the dynamics of the cosmos and the fundamental laws of nature. The persistence of this tension across various probes and over time has increasingly led researchers to interpret it as a call for new physics rather than an artifact of systematics or statistical flukes. However, despite the plethora of attempts, no proposed resolution thus far has proven both observationally and theoretically compelling or widely accepted. Moreover, resolving the $H_0$ tension in a manner that remains consistent with all available data, while avoiding the exacerbation of other less definitive discrepancies such as the $S_8$ tension~\cite{DES:2021wwk,KiDS:2020suj,Heymans:2020gsg,Kilo-DegreeSurvey:2023gfr,Dalal:2023olq,Chen:2024vvk,Kim:2024dmg,DES:2024oud,Harnois-Deraps:2024ucb,Qu:2024sfu,DiValentino:2018gcu,Troster:2019ean,DiValentino:2020vvd,Adil:2023jtu,Akarsu:2024hsu}, remains a formidable challenge. For recent reviews, see~\cite{DiValentino:2021izs,Perivolaropoulos:2021jda,Schoneberg:2021qvd,Shah:2021onj,Abdalla:2022yfr,DiValentino:2022fjm,Akarsu:2024qiq}.

The so-called $\Lambda_{\rm s}$CDM framework~\cite{Akarsu:2019hmw,Akarsu:2021fol,Akarsu:2022typ,Akarsu:2023mfb,Yadav:2024duq,Akarsu:2024eoo,Akarsu:2024qsi} has emerged as one of the most promising and economical extensions of $\Lambda$CDM for addressing major cosmological tensions---including those in $H_0$, $M_{\rm B}$ (SNe Ia absolute magnitude), $S_8$ (growth parameter)---, and the BAO Ly-$\alpha$ anomaly, while yielding an age of the Universe consistent with estimates from the oldest globular clusters and, when allowed to vary, predicting a total neutrino mass and effective number of species in agreement with the Standard Model of particle physics~\cite{Yadav:2024duq}. Notably, this success is achieved by introducing only a single additional parameter beyond the standard $\Lambda$CDM framework. The $\Lambda_{\rm s}$CDM model, originally conjectured in~\cite{Akarsu:2019hmw} based on findings from the graduated dark energy (gDE) model, posits that around redshift $z_\dagger\sim2$, the Universe underwent a period of rapid \textit{mirror} AdS-to-dS (anti-de Sitter-to-de Sitter) transition in vacuum energy---namely, a rapid sign-switching cosmological constant, $\Lambda_{\rm s}$, from negative to positive while preserving its overall magnitude (where ``mirror'' reflects this invariance)---or a similar phenomenon, all while leaving other standard cosmological components, such as cold dark matter, baryons, pre-recombination physics, and inflation, unaltered. From a mathematical/physical perspective, $\Lambda_{\rm s}$CDM is identical to $\Lambda$CDM for $z < z_{\dagger}$, featuring a dS-like CC after the transition, but introduces a minimal modification for $z > z_{\dagger}$ by adopting an AdS-like CC prior to the transition, extending back to the early Universe, including the recombination epoch at $z_{\rm rec} \sim 1100$ and beyond. However, from a phenomenological perspective—i.e., in terms of the Universe’s expansion dynamics and observational signatures—the impact of this modification is effectively confined to redshifts $z \lesssim z_{\dagger}$, with the free parameter $z_\dagger\sim2$ estimated through robust statistical analyses of cosmological data~\cite{Akarsu:2019hmw,Akarsu:2021fol,Akarsu:2022typ,Akarsu:2023mfb,Akarsu:2024eoo,Yadav:2024duq}. Specifically, $\Lambda_{\rm s}$CDM replicates the $H(z)$ of $\Lambda$CDM for $z < z_{\dagger}$—albeit with systematically larger values—introduces localized deformations in $H(z)$ around $z \sim z_{\dagger}$, and becomes nearly indistinguishable from $\Lambda$CDM at higher redshifts ($z \gtrsim 3$). Consequently, from a phenomenological standpoint, $\Lambda_{\rm s}$CDM is a post-recombination/late-time modification to $\Lambda$CDM.

We refer readers (without claiming to be exhaustive) to Refs.~\cite{Vazquez:2012ag,BOSS:2014hwf,Sahni:2002dx,Sahni:2014ooa,Bag:2021cqm,BOSS:2014hhw,DiValentino:2017rcr,Mortsell:2018mfj,Poulin:2018zxs,Capozziello:2018jya,Wang:2018fng,Banihashemi:2018oxo,Dutta:2018vmq,Banihashemi:2018has,Akarsu:2019ygx,Li:2019yem,Visinelli:2019qqu,Ye:2020btb,Perez:2020cwa,Akarsu:2020yqa,DeFelice:2020cpt,Calderon:2020hoc,Ye:2020oix,Paliathanasis:2020sfe,Bonilla:2020wbn,Acquaviva:2021jov,Bernardo:2021cxi,Escamilla:2021uoj,Akarsu:2022lhx,Bernardo:2022pyz,Malekjani:2023ple,Alexandre:2023nmh,Gomez-Valent:2023uof,Medel-Esquivel:2023nov,Tiwari:2023jle,Anchordoqui:2023woo,Anchordoqui:2024gfa,Anchordoqui:2024dqc,Gomez-Valent:2024tdb,Bousis:2024rnb,Wang:2024hwd,Colgain:2024ksa,Yadav:2024duq,Toda:2024ncp,Akarsu:2024nas,Souza:2024qwd,Mukherjee:2025myk,Tyagi:2024cqp,Manoharan:2024thb,Gomez-Valent:2024ejh,Akarsu:2024qsi,Akarsu:2024eoo,Dwivedi:2024okk,Giare:2025pzu,Keeley:2025stf} for further theoretical and observational studies---including model-independent and non-parametric reconstructions---that explore DE models permitting negative energy densities (often consistent with an AdS-like CC) at $z \gtrsim 1.5\text{--}2$, and aimed at addressing major cosmological tensions. Phantom DE models, whose energy densities typically decrease with redshift, though conventionally assumed to remain positive at all redshifts, have been recognized for their potential to alleviate the $H_0$ tension. Among these, the \textit{phantom crossing model}, proposed phenomenologically in~\cite{DiValentino:2020naf} (DMS20~\cite{Adil:2023exv}), stands out. A recent analysis~\cite{Adil:2023exv} reaffirmed its success, but revealed also that its ability to assume negative densities for $z \gtrsim 2$---mimicking an AdS-like CC at sufficiently high redshifts---is central to its effectiveness. Interacting DE (IDE) models~\cite{Kumar:2017dnp,DiValentino:2017iww,Yang:2018uae,Pan:2019gop,Kumar:2019wfs,DiValentino:2019jae,DiValentino:2019ffd,Lucca:2020zjb,Gomez-Valent:2020mqn,Kumar:2021eev,Nunes:2022bhn,Bernui:2023byc,Giare:2024smz,Sabogal:2025mkp} offer another avenue for addressing the $H_0$ tension; yet model-independent reconstructions of the IDE kernel~\cite{Escamilla:2023shf} do not preclude negative DE densities at $z \gtrsim 2$.  Moreover, recent DESI BAO data---analyzed using the CPL parametrization---provided more than $3\sigma$ evidence for dynamical DE~\cite{DESI:2024mwx}. Notably, non-parametric reconstructions of the DE density based on DESI BAO data also suggest the possibility of vanishing or negative DE densities for $z \gtrsim 1.5-2$, a phenomenon similarly observed in pre-DESI BAO data, particularly from SDSS~\cite{Escamilla:2023shf,Sabogal:2024qxs,Escamilla:2024ahl}.

While late-Universe rapid AdS-to-dS (or analogous) transitions in DE, as proposed by the $\Lambda_{\rm s}$CDM framework, were initially viewed as difficult to reconcile with a robust physical mechanism, its notable phenomenological success---despite the model’s simplicity---has spurred deeper theoretical inquiries. Even established frameworks, upon re-examination, have been shown to accommodate such transitions within previously overlooked solution spaces, thereby demanding a fresh perspective on familiar theories. For instance, $\Lambda_{\rm s}$CDM$^+$ proposes a stringy realization of the $\Lambda_{\rm s}$CDM framework; it was shown in~\cite{Anchordoqui:2023woo,Anchordoqui:2024gfa,Anchordoqui:2024dqc} that, despite the AdS swampland conjecture suggesting that an AdS-to-dS transition in the late universe is unlikely due to the arbitrarily large separation between AdS and dS vacua in moduli space, such a transition can nonetheless be realized through the Casimir forces of fields inhabiting the bulk. It was demonstrated in~\cite{Alexandre:2023nmh} that, in certain formulations of general relativity (GR), a sign-switching CC can emerge naturally through an overall sign change of the metric. $\Lambda_{\rm s}$VCDM propels the $\Lambda_{\rm s}$CDM framework to a theoretically complete physical cosmology, providing a fully predictive description of the universe, including the AdS-to-dS transition epoch itself. It was also shown in Refs.~\cite{Akarsu:2024qsi,Akarsu:2024eoo} that the mirror AdS-to-dS transition can be effectively realized within a type-II minimally modified gravity framework called VCDM~\cite{DeFelice:2020eju,DeFelice:2020cpt}, through a particular Lagrangian featuring an auxiliary scalar field with a smoothly sewed two-segmented linear potential. Furthermore, it was shown in Ref.~\cite{Akarsu:2024nas} that the teleparallel $f(T)$ gravity framework, particularly in its exponential infrared formulation~\cite{Awad:2017yod}, which has demonstrated significant potential in addressing the cosmological $H_0$ tension~\cite{Hashim:2020sez,Hashim:2021pkq}, admits previously overlooked solution spaces with promising implications. By relaxing the conventional assumption of a strictly positive effective DE density, it was found that, while ensuring consistency with CMB data, the model accommodates both the well-known phantom behavior and an alternative scenario where the DE density transitions smoothly from negative to positive at redshift $z_\dagger \sim 1.5$. Building on these insights, $f(t)$-$\Lambda_{\rm s}$CDM maps the background dynamics of the $\Lambda_{\rm s}$CDM model into the $f(T)$ gravity framework~\cite{Souza:2024qwd}, further reinforcing its viability as a theoretical underpinning for AdS-to-dS, or similar, transitions in DE in the late universe. Although different realizations of the $\Lambda_{\rm s}$CDM model yield identical background dynamics, a GR-based $\Lambda_{\rm s}$CDM~\cite{Akarsu:2021fol, Akarsu:2022typ, Akarsu:2023mfb}, $\Lambda_{\rm s}$VCDM~\cite{Akarsu:2024qsi,Akarsu:2024eoo}, and $f(T)$-$\Lambda_{\rm s}$CDM would differ in their predictions for linear perturbations, while the string-inspired $\Lambda_{\rm s}$CDM$^+$~\cite{Anchordoqui:2023woo,Anchordoqui:2024gfa,Anchordoqui:2024dqc} predicts a modest excess in the total effective neutrino species, with $\Delta N_{\rm eff} = 0.25$ relative to the standard value of $N_{\rm eff} = 3.044$. Such distinguishing features offer valuable means to compare and ultimately discriminate among these alternative realizations. These examples, which most directly realize the dynamics suggested by the $\Lambda_{\rm s}$CDM framework, are by no means exhaustive. Other models yielding similar behavior include those based on brane world model~\cite{Sahni:2002dx,Sahni:2014ooa,Bag:2021cqm}, energy-momentum log gravity~\cite{Akarsu:2019ygx}, bimetric gravity~\cite{Dwivedi:2024okk}, Horndeski gravity~\cite{Tiwari:2023jle}, holographic DE~\cite{Tyagi:2024cqp}, Granda–Oliveros holographic DE~\cite{Manoharan:2024thb}, composite DE ($w$XCDM)~\cite{Gomez-Valent:2024tdb,Gomez-Valent:2024ejh}, Omnipotent DE~\cite{DiValentino:2020naf,Adil:2023exv}, and scenarios invoking a glitch in the gravitational constant between the super- and sub-horizon regimes, as motivated by the Ho\v{r}ava–Lifshitz proposal or the Einstein-aether framework~\cite{Wen:2023wes}.

In this paper, we instead focus on investigating the possibility of realizing the dynamics suggested by the $\Lambda_{\rm s}$CDM framework, along with its various physical and cosmological implications, primarily at the background level, as well as some of its extensions. Specifically, we explore a minimal extension to the standard $\Lambda$CDM model, wherein the CC is replaced by a minimally coupled canonical scalar field with a suitably chosen potential. This approach, which may have been previously overlooked, allows us to realize the key features of the $\Lambda_{\rm s}$CDM scenario without invoking modifications to GR, in contrast to prior theoretical implementations based on modified gravity frameworks.

The $\Lambda_{\rm s}$CDM framework~\cite{Akarsu:2019hmw,Akarsu:2021fol,Akarsu:2022typ,Akarsu:2023mfb,Yadav:2024duq,Akarsu:2024eoo,Akarsu:2024qsi} essentially generalizes the standard $\Lambda$CDM model by replacing the usual positive CC ($\Lambda$) with a dynamically evolving counterpart ($\Lambda_{\rm s}$) that undergoes a sign-switching transition. This transition can typically be described using sigmoid-like functions, such as the well-known smooth approximation of the signum function, ${\rm sgn}\,x \approx \tanh(kx)$, where $k > 1$ is a constant and $x$ represents either redshift ($z$) or scale ($a$). For instance, consider $\Lambda_{\rm s}(z) = \Lambda_{\rm s0} \,\tanh[\nu(z_{\dagger}-z)]/\tanh[\nu z_\dagger]$, where $\nu > 1$ controls the rapidity of the transition, $\Lambda_{\rm s0} > 0$ is the present-day value of $\Lambda_{\rm s}(z)$, and $z_{\dagger}$ denotes the transition redshift. In the case of a sufficiently rapid transition (e.g., $\nu \gtrsim 10$) occurring at, e.g., $z_{\dagger}=2$, it effectively reduces to positive CC, $\Lambda_{\rm s} \approx \Lambda_{\rm s0}$, for $z \lesssim z_{\dagger}$ and a negative CC, $\Lambda_{\rm s} \approx -\Lambda_{\rm s0}$, for $z \gtrsim z_{\dagger}$. In the limiting case $\nu \to \infty$, the transition becomes instantaneous, recovering the \textit{abrupt} $\Lambda_{\rm s}$CDM model, which extends $\Lambda$CDM by a single additional parameter and is formally described by $\Lambda_{\rm s}(z) \rightarrow \Lambda_{\rm s0} \, {\rm sgn}[z_{\dagger} - z]$ for $\nu \to \infty$, serving as an idealized representation of a rapid mirror AdS-to-dS transition. The abrupt $\Lambda_{\rm s}$CDM scenario, representing the simplest phenomenological realization of the $\Lambda_{\rm s}$CDM framework, has been extensively studied in the literature~\cite{Akarsu:2019hmw,Akarsu:2021fol,Akarsu:2022typ,Akarsu:2023mfb,Yadav:2024duq,Akarsu:2024eoo} under the assumption that its underlying dynamics are governed by GR. However, this abrupt transition model introduces a discontinuity at \( z = z_{\dagger} \), resulting in a type II (sudden) singularity~\cite{Barrow:2004xh}, though a detailed analysis in~\cite{Paraskevas:2024ytz} has demonstrated that its impact on the formation and evolution of cosmic bound structures is negligible. Nevertheless, the presence of this singularity highlights that the abrupt \(\Lambda_{\rm s}\)CDM model should be interpreted as an idealized phenomenological approximation, effectively serving as a proxy for a rapid yet smooth transition. A physically well-defined formulation necessitates a smooth transition, allowing for a detailed investigation of the transition epoch and its consequences, particularly for perturbations. However, implementing a smooth realization introduces additional theoretical challenges that must be thoroughly examined. For instance, a rapid AdS-to-dS transition can lead to a sharp increase in \(\dot{H}\) and potentially induce a transient phase of super-acceleration (\(\dot{H} > 0\)), which may have significant implications for cosmological observables. This is particularly concerning since super-acceleration within GR is often linked to ghost-like instabilities and violations of the weak energy condition. A physically consistent realization of \(\Lambda_{\rm s}\)CDM, therefore, necessitates addressing these issues in a theoretically rigorous manner, extending the framework beyond its purely phenomenological formulation.
 
To rigorously address these challenges, we develop a realization of \(\Lambda_{\rm s}\)CDM~\cite{Akarsu:2019hmw,Akarsu:2021fol,Akarsu:2022typ,Akarsu:2023mfb,Yadav:2024duq,Akarsu:2024eoo,Akarsu:2024qsi} in which the transition is driven dynamically by a minimally coupled scalar field \(\phi\) evolving under a carefully constructed, step-like smooth potential \(V(\phi)\), rather than imposing an ad hoc parametrization of the DE density \(\Lambda_{\rm s}(z)\) in terms of redshift or scale factor. This approach is more rigorous than directly specifying the energy density evolution because it furnishes a complete theoretical framework: it explicitly defines the Einstein–Hilbert action \(S_{\rm EH}\) for gravity, the standard matter action \(S_{\rm m}\), and the scalar field action \(S_\phi\) with a uniquely determined potential that governs dynamical DE. Consequently, this framework enables a detailed investigation of both the Universe’s expansion dynamics and the intrinsic properties of DE, providing a comprehensive basis for a physical cosmology that can be explored over all time and spatial scales—including during the transition epoch. In what follows, we develop this scalar field realization of \(\Lambda_{\rm s}\)CDM, examine its cosmological implications, and explore its extensions, with the aim of achieving a theoretically minimal yet consistent extension of \(\Lambda\)CDM that is fully embedded within the framework of GR.

Scalar fields have long been recognized as a compelling framework for modeling DE, offering a dynamical alternative to a fixed cosmological constant~\cite{Ratra1988,Wetterich1988}. In particular, quintessence models—based on canonical scalar fields—comprise a broad class of theories in which the standard $\Lambda$CDM model emerges as a limiting case when the scalar field is effectively stabilized or “frozen” by cosmic friction in a nearly flat potential~\cite{Sahni2000,Caldwell1998}. Various potential forms have been proposed within this context, including exponential~\cite{Ratra1988}, power-law~\cite{Peebles2003}, and inverse power-law potentials~\cite{Zlatev1999}, each yielding distinct cosmological evolutions and observational signatures. Of particular interest are transition (step-like) potentials, which induce rapid variations in the scalar field dynamics and have been explored as potential mechanisms to address key cosmological tensions—such as those associated with the Hubble constant and the growth of structure~\cite{Verde2019,Perivolaropoulos2021}. Despite their theoretical promise, detailed investigations of these transition potentials remain surprisingly limited. In our formulation within the $\Lambda_{\rm s}$CDM framework, the evolution of the scalar field’s total energy density—and consequently the Hubble parameter $H(z)$—is governed by the interplay between the kinetic term and cosmic friction as the field evolves under a smooth step-like potential. Although the potential naturally induces a smooth, step-like transition in the energy density, the dynamics during the transition epoch are nontrivial; the kinetic term can cause the evolution of the potential and the energy density to deviate significantly from one another. In the nearly flat potential regimes before and after the transition—where the field is slow-rolling and stabilized by cosmic friction—the energy density remains approximately constant, evolving from an initial quasi-stationary state to a similar late-time configuration. Overall, this framework provides a theoretically controlled approach to modeling a smooth AdS-to-dS transition, ensuring a physically robust realization of $\Lambda_{\rm s}$CDM.

In this paper, we conduct a comprehensive study of the background dynamics of smooth transition models driven by scalar fields, with particular emphasis on phantom fields characterized by a negative kinetic term. Although such fields naturally evolve toward the maximum of their potential, they can also give rise to instabilities—such as unbounded energy growth in the absence of an upper bound, Big Rip singularities, and violations of the weak energy condition. We show that adopting a hyperbolic tangent step-like smooth potential circumvents these issues, yielding a physically viable smooth AdS-to-dS transition. By deriving the governing equations for the scalar field in a Friedmann–Lema\^{i}tre–Robertson–Walker background, we analyze the impact of smooth dS/AdS-to-dS transitions on key cosmological observables and compare the predictions of phantom and canonical (quintessence) realizations. Our goal is to establish a solid theoretical foundation for the $\Lambda_{\rm s}$CDM framework and to explore its potential for alleviating the $H_0$ tension through physically motivated dark energy dynamics.

\section{General setup}
\label{sec:model}
We consider a model in which the universe is composed of radiation, dust, represented by pressureless matter---specifically, baryons and cold dark matter (CDM)---along with a minimally interacting (only gravitationally coupled) scalar field $\phi$ with a potential $V(\phi)$, described by the following action:
\begin{equation}
    S_{\phi}=\int {\rm d}^4x\sqrt{-g}\left[\frac{\xi}{2}g^{ik}\partial_{i}\phi\partial_{k}\phi -V(\phi)\right]\, ,
\end{equation}
where $g$ is the determinant of the metric tensor $g_{ik}$, and $g^{ik}$ is its inverse. The indices $i,k=0,1,2,3$ run over the four spacetime coordinates. The parameter $\xi=+1$ corresponds to quintessence (a canonical scalar field with a positive kinetic energy term), while $\xi=-1$ corresponds to a phantom field (a scalar field with a negative kinetic energy term). The energy-momentum tensor and equation of motion for this scalar field are given by, respectively:
\begin{align}
T_k^i(\phi)=\xi g^{i l}\partial_l\phi\partial_k\phi-\delta^i_k\left(\frac{\xi}{2}g^{l s}\partial_l\phi\partial_s\phi- V(\phi)\right)\ ,\\
 \frac{\xi}{\sqrt{-g}}\partial_i\left(\sqrt{-g}g^{ik}\partial_k\phi\right)+\frac{{\rm d}V}{{\rm d}\phi}=0\,.
\end{align}

Throughout this work, we assume the dynamics of the universe is governed according to the general relativity and describe the background geometry of the universe using the spatially maximally symmetric spacetime, viz., Friedmann-Lema\^{\i}tre-Robertson-Walker metric, with flat spatial sections:
\begin{equation}
   {\rm d}s^2=c^2{\rm d}t^2-a^2\delta_{\alpha\beta}{\rm d}x^\alpha {\rm d}x^\beta \, , 
\end{equation}
where $c$ is the speed of light, $a=a(t)$ is the scale factor, $t$ is cosmic time, and $x^{\alpha}$ (with $\alpha =1,2,3$) are the comoving spatial coordinates. Assuming the components of the universe are minimally interacting (interacting only gravitationaly) the energy density of pressureless-matter~(m) and of radiation~(r) evolve as
\begin{equation}
    \varepsilon_{\rm m} = \varepsilon_{\rm m0} /a^3, \qquad  \varepsilon_{\rm r} = \varepsilon_{\rm r0} /a^4\, ,
\label{eq:matter_and_radiation_density}
\end{equation}
where $\varepsilon_{\rm m0}=\rm{const}$ and $\varepsilon_{\rm r0}=\rm{const}$ are the present-day energy densities, and we adopt the normalization $a_0 =1$ for the scale factor at the present time. The evolution of the universe is governed by the Friedmann equations:
\begin{align}
   \frac{3H^2}{c^2} =\kappa \left(\varepsilon_{\rm m}+\varepsilon_{\rm r}+\varepsilon_\phi\right) \,, \label{fried1}  \\
   \frac{2\dot H + 3H^2}{c^2}= -\kappa\, (p_{\rm r} + p_{\phi})\, ,\label{fried2}
\end{align}
where $\kappa \equiv 8\pi G_N/c^4$, with $G_N$ being the Newtonian gravitational constant. The scalar field $\phi$ has units of $\kappa^{-1/2}$, corresponding to the dimension of mass in natural units where $\hbar =c=1$. Consequently, the combination $\phi\sqrt{\kappa}$ is dimensionless.
For the background, real-valued scalar field $\phi=\phi(t)$, the non-zero components of the energy-momentum tensor, which define the scalar field's energy density $\varepsilon_{\phi}$ and pressure $p_{\phi}$, which is isotropic, are given by:
\begin{align}
    \varepsilon_{\phi}\equiv{T^0_0}(\phi)=\frac{\xi}{2c^2}(\dot\phi)^2+V(\phi) \, , \label{energyphi} \\
    p_{\phi}\equiv-T^{\alpha}_{\alpha}(\phi) =\frac{\xi}{2c^2}(\dot\phi)^2-V(\phi) , \label{pressurephi}
\end{align}
where the dot denotes differentiation with respect to the cosmic time $t$. The Klein-Gordon equation, i.e., the equation of motion, for the scalar field is given by:
\begin{equation}
  \ddot\phi+3H\dot\phi + \xi c^2\frac{dV}{d\phi}=0\, ,
\label{fielddyneq}
\end{equation}
where $H=\dot a/a$ is the Hubble parameter. The equation of state (EoS) parameter of the scalar field is then:
\begin{equation}
  \omega_\phi=\frac{p_\phi}{\varepsilon_\phi}=\frac{\xi X-V(\phi)}{\xi X+V(\phi)},\quad X\equiv \frac{1}{2c^2}(\dot\phi)^2\,, 
\end{equation}
where $X>0$ as we consider (real valued) scalar field $\phi=\phi(t)$. It is straightforward to see that the EoS parameter $\omega_\phi$ can lie on either side of the phantom divide line $-1$. In particular, depending on whether $\xi=+1$ or $-1$ and whether $\varepsilon_\phi$ is positive ({\it p}-) or negative ({\it n}-), one obtains four branches:
\begin{itemize}[nosep,wide]
    \item $\xi = +1$ (Quintessence):
    \begin{itemize}
        \item (\textit{p-quintessence}) $\omega_\phi > -1$ if $V > -X$, indicating a positive energy density ($\varepsilon_{\phi} > 0$). This corresponds to the standard quintessence scenario.
        \item (\textit{n-quintessence}) $\omega_\phi < -1$ if $V < -X$, implying that the quintessence field attains negative energy density ($\varepsilon_{\phi} < 0$).
    \end{itemize}

    \item $\xi = -1$ (Phantom):
    \begin{itemize}
        \item (\textit{{\it p}-phantom}) $\omega_\phi < -1$ if $V > X$, indicating a positive energy density ($\varepsilon_{\phi} > 0$). This corresponds to the familiar phantom scenario.
        \item (\textit{{\it n}-phantom}) $\omega_\phi > -1$ if $V < X$, implying that the phantom field attains negative energy density ($\varepsilon_{\phi} < 0$).
    \end{itemize}
\end{itemize}

It is crucial to emphasize that the scalar field’s speed of sound squared, $c_s^2$, always remains equal to $c^2$, regardless of $\xi$, the form of $V(\phi)$, or even the sign of the scalar field’s energy density $\varepsilon_{\phi}$. This result follows from the Lagrangian’s linear dependence on the kinetic term $X$. Specifically, the pressure $p_{\phi}$ and energy density $\varepsilon_{\phi}$, as given in Eqs.~\eqref{energyphi} and \eqref{pressurephi}, satisfy:
\begin{equation}
    c_s^2 
    \;=\;
    \frac{\partial p_{\phi}}{\partial \varepsilon_{\phi}}\,c^2
    \;=\;
    \frac{p_{\phi, X}}{\varepsilon_{\phi, X}}\,c^2
    \;=\;
    \frac{\xi}{\xi}\,c^2
    \;=\;
    c^2,
\end{equation}
where $p_{\phi, X}$ and $\varepsilon_{\phi, X}$ denote the derivatives of $p_{\phi}$ and $\varepsilon_{\phi}$ with respect to $X$. Because the factor $\xi$ cancels out, the characteristic speed of perturbations in both quintessence and phantom fields is always the speed of light, $c$, independently of the sign of $\varepsilon_{\phi}$. Consequently, scalar-field perturbations always propagate at the speed of light.

We note that the scalar field is inherently confined to be either quintessence or phantom, depending on whether $\xi = +1$ or $\xi = -1$, respectively. Observational evidence indicates that the present-day dark energy density is positive, $\varepsilon_\phi(t = t_0) > 0$, suggesting that the scalar field could have remained entirely on the p-quintessence or {\it p}-phantom branch throughout cosmic history. However, this need not necessarily be the case. It is also conceivable that the scalar field initially exhibited a negative energy density, $\varepsilon_\phi < 0$, before transitioning to positive values at later times. This possibility invites consideration of transitions such as $n$-phantom to $p$-phantom or $n$-quintessence to $p$-quintessence at some point during the evolution of the universe. Importantly, transitions between distinct classes, such as $n$-phantom to $p$-quintessence or $n$-quintessence to $p$-phantom, are fundamentally forbidden, as the scalar field is strictly confined to remain either quintessence or phantom. This restriction arises directly from the inherent structure of the field's dynamics, dictated by the fixed sign of $\xi$, which governs the behavior of its kinetic and potential terms.

A succinct way to demonstrate  the contrasting behaviors of quintessence $(\xi=+1)$ and phantom $(\xi=-1)$ is to employ the Klein--Gordon equation~\eqref{fielddyneq}, where $\ddot{\phi}$ is the field acceleration, $-\,3H\,\dot{\phi}$ is the friction-like term, and $-\,\xi\,c^2\,\tfrac{dV}{d\phi}$ acts as an ``effective force.'' For quintessence $(\xi=+1)$, this force is $-\,c^2\,\tfrac{dV}{d\phi}$, directing $\phi$ to roll downhill toward a minimum of $V(\phi)$. Consequently, if the total scalar-field energy density $\varepsilon_{\phi}=X+V(\phi)$ is already negative, rolling down only pushes it more negative. By contrast, for phantom fields $(\xi=-1)$, the sign of the potential gradient flips, so the field is effectively propelled uphill, allowing $\varepsilon_{\phi}$ to increase in spite of cosmic expansion. A particularly illustrative scenario arises if the phantom field potential is step-shaped,  such as a hyperbolic tangent form, featuring two gently positively tilted plateaus at different energy levels connected by a relatively much steeply increasing sigmoid region. Initially, the phantom field may reside on a negative-valued (AdS-like) plateau of $V(\phi)$ with $X \ll \lvert V(\phi)\rvert$. During early epochs---e.g.\ matter domination---the friction $-\,3H\,\dot{\phi}$ can initially be substantial due to large $H$ values, which tends to keep slow/slow down the field's motion, maintaining $\varepsilon_{\phi}\approx V(\phi)<0$. However, as the universe expands and $H$ diminishes (as expected when the usual cosmological fluids are in act), the uphill force $+\,c^2\tfrac{dV}{d\phi}$ becomes dominant, causing $\phi$ to climb the potential with increasing speed. When the scalar field reaches the nearly flat plateau at higher level (dS-like), passing through the steeply inclined region of the potential, the gradient $\tfrac{dV}{d\phi}$ becomes very small again, allowing the friction term to dominate once more. Consequently, the field slows and settles near this dS-like plateau, where $\varepsilon_{\phi} \approx V(\phi) > 0$. In this final state, the scalar field behaves like a positive CC, contributing positively to $H(z)$ in the late universe.

To further analyze the feasibility of such transitions, we invoke the continuity equation for the scalar field:
\begin{equation}\label{eq:continuity}
\dot{\varepsilon}_{\phi} \;=\; -3H\,(\varepsilon_{\phi}+p_{\phi}\bigr)=-3H\,\varepsilon_{\phi}\,\bigl(1+\omega_{\phi}\bigr)\, ,
\end{equation}
where $H>0$ is the Hubble parameter for an expanding universe.  The sign of $\dot{\varepsilon}_{\phi}$ is determined by the product $\varepsilon_{\phi}\,(1 + \omega_{\phi})$. Thus, if $\varepsilon_{\phi}\,(1+\omega_{\phi}) > 0$, then $\dot{\varepsilon}_{\phi} < 0$, whereas if $\varepsilon_{\phi}\,(1+\omega_{\phi})<0$, we have $\dot{\varepsilon}_{\phi}>0$. 

\textit{{\it n}-phantom to {\it p}-phantom Crossing}: In the $n$-phantom branch, characterized by $\varepsilon_{\phi} < 0$ and $\omega_{\phi} > -1$, the continuity equation implies that $\dot{\varepsilon}_{\phi} > 0$, indicating that the energy density increases over time (from more negative to less negative) as the universe expands. Consequently, $\varepsilon_{\phi}$ can smoothly approach zero from below, crossing $\varepsilon_{\phi} = 0$ in finite time without requiring any singular behavior. Both $p_\phi$ and $\varepsilon_\phi$ evolve smoothly; the divergence of $\omega_{\phi}$ occurs solely because $\varepsilon_\phi$ crosses zero (implying $X = V$), while $p_\phi = -2X = -2V < 0$ remains finite, and $\dot{\varepsilon}_{\phi} \to -3H p_\phi = 6HX$ as $\varepsilon_{\phi} \to 0$. To elucidate the behavior of $\omega_{\phi}$ near $\varepsilon_{\phi} = 0$ (i.e., $X = V$), we parametrize the potential as $V = X \pm \delta$, where $\delta \geq 0$. For $V < X$ (i.e., $V = X - \delta$), the EoS parameter becomes $\omega_{\phi} = -1 + \frac{2X}{\delta}$, which diverges to $+\infty$ as $\delta \to 0^{+}$ (i.e., $\varepsilon_{\phi} \to 0^{-}$). Conversely, for $V > X$ (i.e., $V = X + \delta$), we find $\omega_{\phi} = -1 - \frac{2X}{\delta}$, which diverges to $-\infty$ as $\delta \to 0^{+}$ (i.e., $\varepsilon_{\phi} \to 0^{+}$). Away from the transition region, where $X \ll |V|$, the ratio $\frac{X}{\delta} \ll 1$, ensuring that $\omega_{\phi}$ remains close to $-1$ (mimicking a CC). Accordingly, $\omega_{\phi}$ resides above $-1$ before the transition, exhibits a safe singularity during the $\varepsilon_{\phi} = 0$ crossing, and then asymptotically approaches $-1$ from below (see Ref.~\cite{Ozulker:2022slu} examining singularities in EoS parameter). Importantly, as established earlier, the sound speed squared $c_s^2$ always remains equal to $c^2$, ensuring that scalar-field perturbations propagate without any physical anomalies or pathological behavior.

\textit{Impossibility of n-Quintessence to p-Quintessence transition under Smooth Expansion}: The situation is dramatically different for {\it n}-quintessence, characterized by $\varepsilon_{\phi} < 0$ and $\omega_{\phi} < -1$. In this regime, since $\varepsilon_{\phi} < 0$ and $1 + \omega_{\phi} < 0$, the product $\varepsilon_{\phi}\,(1 + \omega_{\phi}) > 0$. Consequently, the continuity equation yields $\dot{\varepsilon}_{\phi}= -6HX < 0$, indicating that the negative energy density becomes increasingly negative as the universe expands. This prevents any zero-crossing of $\varepsilon_{\phi}$. In fact, transitioning from $\varepsilon_{\phi} < 0$ to $\varepsilon_{\phi} > 0$ under the strict condition $\omega_{\phi} < -1$ would require $\varepsilon_{\phi}$ to diverge to $-\infty$ in finite time and then re-emerge from $+\infty$, constituting a “bounce via infinity” that represents a genuine singularity or breakdown of the standard Friedmann evolution, typically considered unphysical. Therefore, under normal continuity of solutions and without violating $\omega_{\phi} < -1$, a transition from {\it n}-quintessence ($\varepsilon_{\phi} < 0$) to {\it p}-quintessence ($\varepsilon_{\phi} > 0$) is forbidden in an expanding universe. This contrast between the phantom and quintessence cases underscores the subtlety of sign changes in scalar-field energy densities. The precise mechanism that allows {\it n}-phantom crossing—where $\omega_{\phi} > -1$—is exactly what fails in the {\it n}-quintessence scenario, where $\omega_{\phi} < -1$ drives the energy density to become increasingly negative.

These results establish a no-go theorem for transitions from {\it n}-quintessence to {\it p}-quintessence under smooth cosmic expansion, while permitting an elegant, non-singular zero-crossing from {\it n}-phantom to {\it p}-phantom. Essentially, the sign of $1 + \omega_{\phi}$ dictates the direction of energy-density evolution: an {\it n}-phantom field can ascend from negative to positive energy density, whereas an {\it n}-quintessence field is compelled to descend to increasingly negative energy densities. This fundamental distinction ensures that only {\it n}-phantom fields can undergo smooth transitions to positive energy states without encountering singularities, highlighting the unique dynamical properties of phantom scalar fields in cosmological evolution.

In the preceding discussion, we have qualitatively explored the possibility of AdS-to-dS transitions in the context of a phantom field, emphasizing their dynamical plausibility and connection to late-time cosmic acceleration. A deeper understanding of these transitions requires specifying the precise form of the potential $V(\phi)$, which fundamentally governs the scalar field's behavior. These models stand as intriguing alternatives to the standard $\Lambda$CDM paradigm, offering a framework that can address persistent challenges in cosmology, such as the $H_0$ tension. The following sections are dedicated to a detailed exploration of these models, focusing on their ability to reconcile key observational discrepancies and refine our understanding of late-time cosmic evolution.

\section{Dynamical dark energy with ${\rm AdS} \leftrightarrow {\rm dS}$ transitions}
\label{sec:AdS-to-dS}

 To explore the possibilities for AdS $\leftrightarrow$ dS transitions, we consider the scalar field potential of the form:
\begin{equation}
    V(\phi)= \dfrac{\Lambda(\xi_1+1)}{2} -\dfrac{\Lambda(\xi_1-1)}{2} \tanh\left[\sqrt{\kappa}\nu\left(\phi-\phi_{\rm c}\right)\right],
\label{eqn:gen_pot}
\end{equation}
where $\Lambda > 0$ is a characteristic energy scale (e.g., the present-day value of the potential), $\xi_1$ is a dimensionless parameter controlling the initial vacuum energy value, $\nu$ dictates the rapidity of the transition, and $\phi_{\rm c}$ corresponds to the inflection point, marking the midpoint of the transition where the hyperbolic tangent changes sign. This potential exhibits the following asymptotic behaviors: $ V(\phi \to -\infty) \to \Lambda\xi_1$ and $ V(\phi \to +\infty) \to \Lambda$. In the special case where $\xi_1 = 1$, the potential simplifies to $V(\phi) = \Lambda$, thereby recovering the standard $\Lambda$CDM model. Here, the scalar field becomes effectively frozen due to the strong cosmological damping induced by the term $3H\dot{\phi}$ in the scalar field equation of motion~\cref{fielddyneq}, resulting in behavior characteristic of a positive CC $\Lambda$.

 We consider scenarios in which the initial value of the scalar field is chosen to be prior to the transition, such that $\phi_{\rm in}<\phi_{\rm c}$. The nature of the transition---whether AdS-to-dS or dS-to-dS---depends on the value of the parameter $\xi_1$, allowing for different outcomes for both phantom ($\xi = -1$) and quintessence ($\xi = +1$) fields. Specific examples of these transitions are discussed in detail in subsequent sections. In all cases, the parameter $\nu$ dictates the rapidity of the transition, controlling the steepness of the potential as the field evolves through the transition phase. To analyze these transitions, it is useful to describe the dynamics of the scalar field and the universe in terms of the redshift, defined as $z = a_0/a-1$. By adopting the gauge $a_0=1$ for the scale factor at the present time, this simplifies to $z=1/a-1$, from which we can derive that the time derivative transforms as  ${\rm d}/{\rm d}t = -(1+z)H(z) {\rm d}/{\rm d}z$. Introducing the dimensionless parameters:
\begin{equation}
\begin{aligned}
    \Omega_{\rm m0}&\equiv \frac{\varepsilon_{\rm m0}}{\varepsilon_{\rm{cr}0}}\, ,\; \Omega_{\rm r0}\equiv \frac{\varepsilon_{\rm r0}}{\varepsilon_{\rm{cr}0}}\, ,\; \Omega_{\Lambda 0}\equiv\frac{\Lambda}{\varepsilon_{\rm{cr}0}}\, ,\; \\[1mm]
     \widetilde\Omega_{\phi}(z) &\equiv \frac{\varepsilon_{\phi}}{\varepsilon_{\rm{cr}0}}\, ,\; \widetilde h(z) \equiv \frac{H}{H_0}\, , \; \tilde\phi(z) \equiv \sqrt{\kappa}\phi \, , \\[1mm]
    \widetilde\Omega_{K}(z) &\equiv \frac{\xi}{2c^2}\frac{(\dot\phi)^2}{\varepsilon_{\rm{cr}0}}=
    \frac{\xi}{6}(1+z)^2{\widetilde h}^2\left(\frac{{\rm d}\widetilde \phi}{{\rm d}z}\right)^{\!\!2} \, , \\[1mm]
    \widetilde\Omega_{V}(z)&\equiv \frac{V}{\varepsilon_{\rm {cr}0}} = \Omega_{\Lambda 0} \bigg\{ \frac{\xi_1+1}{2} \\[1mm]
    & \hspace{15mm} - \frac{\xi_1-1}{2}\tanh\left[\nu\left(\tilde\phi-\tilde\phi_{\rm c}\right)\right]\bigg\}\, ,
\label{parameters}
\end{aligned}
\end{equation}
where $\varepsilon_{\rm{cr}0}=3 H_0^2/\kappa c^2$ is the critical energy density and $H_0=H(z=0)$ is the Hubble constant,
we rewrite Eqs.~\eqref{eq:matter_and_radiation_density} to \eqref{fielddyneq} in the following dimensionless forms
\begin{align}
    \widetilde\Omega_{\rm m}(z) = \Omega_{\rm m0} (1+z)^3 \, , \quad \widetilde\Omega_{\rm r}(z) = \Omega_{\rm r0} (1+z)^4 \, ,
\end{align}
\begin{align}
    \widetilde\Omega_{\phi} = \widetilde\Omega_K + \widetilde\Omega_V \, ,
\label{energy_phi_dless}
\end{align}
\begin{align}
    \frac{p_{\phi}}{\varepsilon_{\rm{cr}0}} = \widetilde\Omega_K - \widetilde\Omega_V \, ,
\label{pressure phi dless}
\end{align}
\begin{equation}
\begin{aligned}
    &(1+z)^2{\widetilde h}^2\frac{{\rm d}^2\tilde\phi}{{\rm d}z^2} + (1+z)^2{\widetilde h}\frac{{\rm d}\widetilde h}{{\rm d}z}\frac{{\rm d}\widetilde \phi}{{\rm d}z}\\[2mm]
    & \hspace{25mm} - 2(1+z){\widetilde h}^2\frac{{\rm d}\widetilde \phi}{{\rm d}z}+3\xi\frac{{\rm d}\widetilde\Omega_V}{{\rm d}\tilde\phi}=0\, ,
    \end{aligned}
\label{eq_phi_dless}
\end{equation}
\begin{align}
    \widetilde h^2 = \widetilde\Omega_{\rm m} + \widetilde\Omega_{\rm r} + \widetilde\Omega_{\phi}\, ,
\label{friedmann_1_dless}
\end{align}
and
\begin{align}
    2(1+z){\widetilde h}\frac{{\rm d}\widetilde h}{{\rm d}z} = & \: 3\widetilde\Omega_{\rm m} + 4\widetilde\Omega_{\rm r} + 6\widetilde\Omega_K \; .
\label{friedmann_2_dless}
\end{align}
It is noteworthy that the dimensionless parameters denoted with a tilde $\widetilde\Omega_{\rm m,\rm r}$ are connected to the standard density parameters $\Omega_{\rm m,\rm r} \equiv \varepsilon_{\rm m,\rm r}/\varepsilon_{\rm{cr}}$, where $\varepsilon_{\rm{cr}}(z) = 3 H^2 / (\kappa c^2)$ is the critical energy density at redshift $z$, as follows: 
\begin{equation} 
\widetilde\Omega_{\rm m,\rm r} = \Omega_{\rm m,\rm r} \widetilde{h}^2\, . \label{3.9} 
\end{equation} 
Similarly, this relation applies to $\widetilde\Omega_{\phi}$, $\widetilde\Omega_{K}$, and $\widetilde\Omega_{V}$. We define the EoS parameter for the scalar field as
\begin{align}
    \omega_{\phi}=\frac{p_{\phi}}{\varepsilon_{\phi}} = \dfrac{\widetilde\Omega_{K}-\widetilde\Omega_V}{\widetilde\Omega_{K}+\widetilde\Omega_V} \, ,
\label{eq state phi}
\end{align}
and for total content of the universe as
\begin{align}
    \omega_{\rm{tot}}=\dfrac{p_{\rm{tot}}}{\varepsilon_{\rm{tot}}} \, ,
\label{eq state tot}
\end{align}
where
\begin{equation}
\begin{aligned}
    \dfrac{p_{\rm{tot}}}{\varepsilon_{\rm{cr}0}} = \dfrac{p_{\rm{m}} + p_{\rm{r}} + p_{\phi}}{\varepsilon_{\rm{cr}0}} &= \dfrac{p_{\rm{r}} + p_{\phi}}{\varepsilon_{\rm{cr}0}} 
    = \dfrac{1}{3}\widetilde\Omega_{\rm r} + \widetilde\Omega_K - \widetilde\Omega_V\, ,
\label{pressure_total}
\end{aligned}
\end{equation}
and as a dimensionless counterpart of $\varepsilon_{\rm{tot}}$ we can define
\begin{equation}
\begin{aligned}
    \widetilde\Omega_{\rm tot}(z) = \dfrac{\varepsilon_{\rm{tot}}}{\varepsilon_{\rm{cr}0}} &= \dfrac{\varepsilon_{\rm{m}} + \varepsilon_{\rm{r}} + \varepsilon_{\phi}}{\varepsilon_{\rm{cr}0}} 
    &=\widetilde\Omega_{\rm m} + \widetilde\Omega_{\rm r} + \widetilde\Omega_K +\widetilde\Omega_V \,.
\label{energy tot dless}
\end{aligned}
\end{equation}

It is evident that the positivity of $\widetilde\Omega_{\rm tot}$ ensures the positivity of the total energy density $\varepsilon_{\rm{tot}}>0$. This condition is fundamental for maintaining the physical viability of the model, as negative total energy densities could lead to unphysical scenarios. From the perspective of energy conditions, particularly the null energy condition (NEC) applied to the total content of the universe (implying $p_{\rm tot} + \varepsilon_{\rm tot}>0$), it is insightful to consider the dimensionless combination:
\begin{equation}
\begin{aligned}
    \dfrac{p_{\rm tot} + \varepsilon_{\rm tot}}{\varepsilon_{\rm{cr}0}}= \dfrac{p_{\rm r} + p_{\phi} + \varepsilon_{\rm tot}}{\varepsilon_{\rm{cr}0}}
    = \widetilde\Omega_{\rm m} + \dfrac{4}{3}\widetilde\Omega_{\rm r} + 2\widetilde\Omega_K \, .
\end{aligned}
\label{p + e tot dless}
\end{equation}

The following two equations are fundamental for analyzing the dynamics of the universe. The first equation governs the rate of change of the Hubble parameter:
\begin{equation}
\begin{aligned}
    \dfrac{\dot H}{H_0^2} = -(1+z)\widetilde{h}\dfrac{{\rm d}\widetilde{h}}{{\rm d}z} = -\dfrac{1}{2} \big( 3\widetilde\Omega_{\rm m} + 4\widetilde\Omega_{\rm r} + 6\widetilde\Omega_K \big) \, .
\end{aligned}
\label{H dot dless}
\end{equation}
This equation is derived by differentiating the Friedmann equation and encapsulates how the expansion rate changes due to contributions from matter, radiation, and the kinetic energy of the scalar field. Note that the potential term does not contribute to this equation. The second equation defines the dimensionless deceleration parameter $ q $, which characterizes the acceleration or deceleration of cosmic expansion:
\begin{equation}
\begin{aligned}
    q = -1 + \dfrac{{\rm d}}{{\rm d}t}\frac{1}{H} &= -1 + \dfrac{(1+z)\widetilde h}{{\widetilde h}^2}\frac{{\rm d}\widetilde h}{{\rm d}z}\\[2mm]
    &= \dfrac{\dfrac{1}{2} \widetilde\Omega_{\rm m} + \widetilde\Omega_{\rm r} + 2\widetilde\Omega_K - \widetilde\Omega_V}{\widetilde\Omega_{\rm m} + \widetilde\Omega_{\rm r} + \widetilde\Omega_K + \widetilde\Omega_V} \, .
\end{aligned}
\label{dec par }
\end{equation}
The deceleration parameter $ q < 0 $ signifies an accelerating universe, while $ q > 0 $ indicates deceleration. This parameter is crucial for understanding the transitions between different cosmological epochs and the influence of dark energy on the universe's expansion.

In the subsequent sections, we will explore the most compelling scenarios for both quintessence and phantom scalar fields, examining their roles in AdS-to-dS and dS-to-dS transitions and their implications for cosmological observations.


\section{Mirror ${\rm AdS} \rightarrow {\rm dS}$ transition: $\Lambda_{\rm s}$CDM model}
\label{sec:AdS-to-dS}

In this section, we explore the AdS-to-dS transition driven by a phantom scalar field ($ \xi = -1 $). It is well known that such a field violates the null energy condition (NEC), as can straightforwardly seen from~\cref{energyphi,pressurephi}, where the inertial mass density, defined as $ \varrho_{\phi} \equiv \varepsilon_{\phi} + p_{\phi} $, satisfies $ \varrho_{\phi} = -\frac{1}{c^2}(\dot{\phi})^2 < 0 $. Furthermore, the energy density of the scalar field can become negative, i.e., $ \varepsilon_{\phi} < 0 $. However, when additional matter components---such as cold dark matter (CDM)---are present, the total inertial mass density is given by $ \varrho_{\rm tot} = \varrho_{\phi} + \varrho_{\rm m} $, where $ \varrho_{\rm m} = \varepsilon_{\rm m} $ for dust. In this case, the total inertial mass density can remain non-negative at all times, ensuring $ \varrho_{\rm tot} \geq 0 $ and $ \varepsilon_{\rm tot} \geq 0 $. This balance is essential in our model, maintaining physical viability within the context of the phantom field’s dynamics. 

For phantom models with positive energy density, the dynamics typically lead to a big rip singularity, where the scale factor and energy density diverge within a finite time horizon~\cite{Caldwell:1999ew, Starobinsky:1999yw, Caldwell:2003vq, Johri:2003rh, Novosyadlyj:2012vd}. However, if the asymptotic geometry is de Sitter, the universe avoids the big rip scenario~\cite{Bouhmadi-Lopez:2004mpi}. It is crucial to recognize that the negative kinetic term intrinsic to phantom fields inherently destabilizes the model unless the potential is suitably bounded from above. Consequently, both the onset of a big rip and the instability associated with the phantom's negative kinetic energy can be mitigated through a careful selection of the potential. The potential introduced in~\cref{eqn:gen_pot} is specifically designed to address these issues, ensuring the model remains stable while preventing singular behavior.

We focus on the case of a mirror AdS-to-dS transition realized when $\xi_1=-1$. In this scenario, the potential in~\cref{eqn:gen_pot} takes the form
\begin{equation}
V(\tilde\phi)=\Lambda\tanh\left[\nu(\tilde\phi -\tilde\phi_{\rm c})\right],\quad \Lambda>0\, ,
 \label{potential_lscdm}
\end{equation}
with $ V(\tilde\phi) \approx -\Lambda$ for $\nu(\tilde\phi -\tilde\phi_{\rm c})\ll0$ and $ V(\tilde\phi) \approx \Lambda$ for $\nu(\tilde\phi -\tilde\phi_{\rm c})\gg0$. The dimensionless counterpart, $\widetilde\Omega_V$, is provided in~\cref{parameters} for $\xi_1=-1$. This form of potential is particularly interesting, as the mirror AdS-to-dS transition offers a natural explanation for the $\Lambda_{\rm s}$CDM model. A similar potential was proposed in~\cite{Akarsu:2024qsi} to describe a smooth AdS-to-dS within the $\Lambda_{\rm s}$CDM paradigm, realized in the VCDM framework~\cite{DeFelice:2020eju,DeFelice:2020cpt}---a type-II minimally modified gravity. Here, we demonstrate that such a transition can also be realized within GR using a phantom field.

\begin{table}[!t]
\label{table:I}
\renewcommand{\arraystretch}{1.2}
\centering
\caption{The values of parameters $z_{\rm t}$ $z_\dagger$, and $\Delta z\equiv z_{\rm t}-z_\dagger$ depending on the choice of parameters $H_0$ and $\nu$ in the case of initial conditions $\tilde\phi_{\rm{in}} =\tilde\phi'_{\rm{in}} =0$. 
Parameter $\Omega_{\rm m0}$ is calculated by~\cref{eq:Omega_M0}. The case $\nu=15$ (exist in~\cref{fig:H0_zdag}) is not given, as it gives results very similar to those for $\nu=10$.}
\begin{tabular}{C{0.8cm}|C{1.4cm}|C{0.9cm}|C{1.4cm}||C{1cm}|C{1cm}|C{1cm}}
\hline
$H_0$ & $\Omega_{\rm m0}$ & $\nu$ & $\tilde\phi_{\rm{c}}$ & $z_{\rm t}$ & $z_\dagger$ & $\Delta z$ \\ \hline

75 & 0.2545 & \multirow{5}{*}{10} & 0.1269 & 1.871 & 1.545 & $0.326$ \\ 
73 & 0.2687 & & 0.1106 & 2.104 & 1.744 & $0.360$ \\
71 & 0.2840 & & 0.0861 & 2.547 & 2.122 & $0.425$ \\ 
69 & 0.3007 & & 0.0314 & 4.382 & 3.707 & $0.675$ \\
68 & - & & - & - & - & -  \\ \hline

75 & 0.2545 & \multirow{6}{*}{100} & 0.0356 & 1.919 & 1.609 & 0.310 \\ 
73 & 0.2686 & & 0.0342 & 2.127 & 1.796 & 0.331 \\
71 & 0.2840 & & 0.0321 & 2.476 & 2.109 & 0.367 \\ 
69 & 0.3007 & & 0.0287 & 3.251 & 2.807 & 0.444\\
68 & 0.3096 & & 0.0250 & 4.347 & 3.791 & 0.556\\ 
67 & - & & - & - & - & -\\ \hline

75 & 0.2545 & \multirow{6}{*}{1000} & 0.00585 & 1.929 & 1.610 & 0.319  \\ 
73 & 0.2686 & & 0.00571 & 2.139 & 1.798 & 0.341 \\
71 & 0.2840 & & 0.00550 & 2.491 & 2.112 & 0.379 \\ 
69 & 0.3007 & & 0.00515 & 3.278 & 2.813 & 0.465 \\ 
68 & 0.3096 & & 0.00478 & 4.392 & 3.809 & 0.583 \\
67 & - & & - & - & - & - \\ \hline
\end{tabular}\label{tab1}
\end{table}

\begin{figure*}[ht!]
\centering
    \begin{tabular}{@{}c@{}}\hspace{-3mm}
	\includegraphics[width=0.75\linewidth]{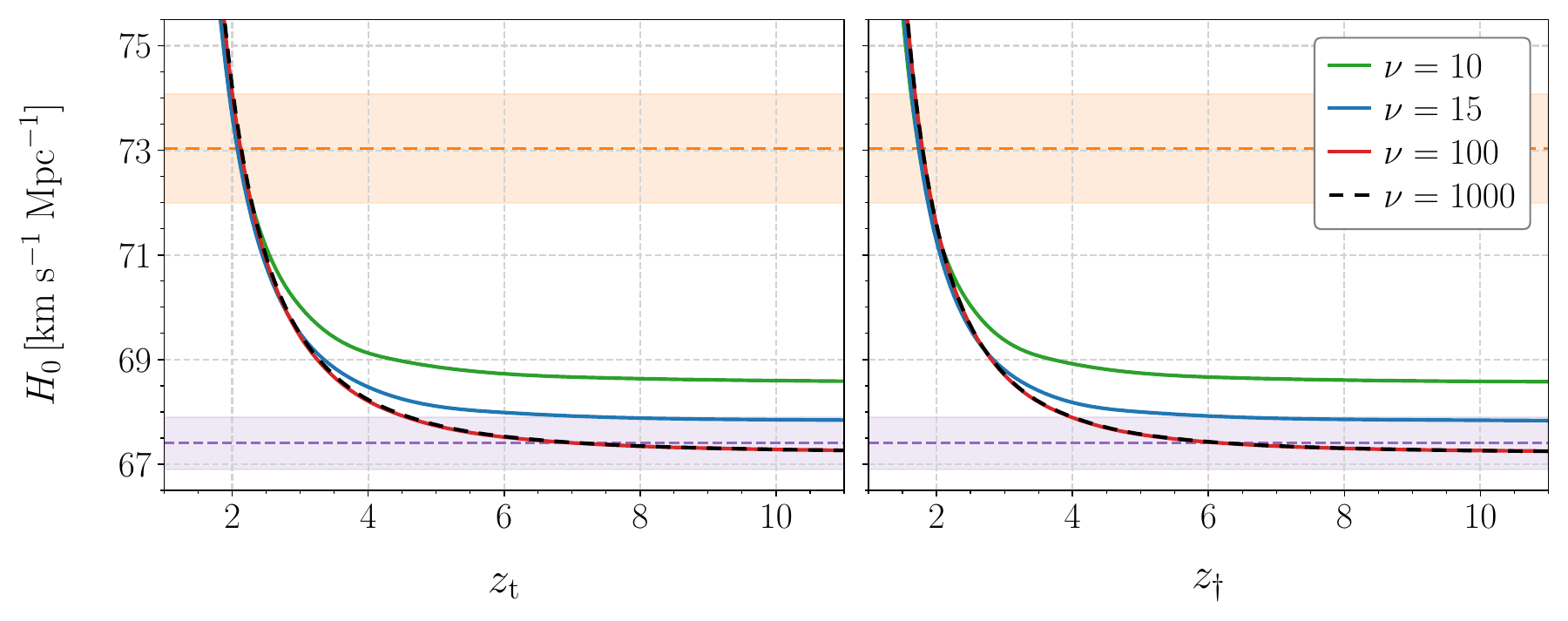}
    \end{tabular}
    \vspace{-3mm}
    \caption{Dependence of $H_0$ on $z_{\rm t}$ and $z_\dagger$ for different values of the rapidity parameter $\nu$.
    The horizontal red line and shaded region correspond to the SH0ES result $H_0=73.04 \pm 1.04~{\rm km\, s^{-1}\, Mpc^{-1}}$~\cite{Riess:2021jrx} whereas the horizontal purple line and shaded region indicate the Planck data $H_0=67.4 \pm 0.5~{\rm km\, s^{-1}\, Mpc^{-1}}$~\cite{Planck:2018vyg}. 
    We see that for all selected values of the rapidity parameter $\nu$ the present day value $H_0$ can reach  the SH0ES result $73.04~{\rm km\, s^{-1}\, Mpc^{-1}}$ at $z_{\rm t}\approx 2.1$.  By contrast, the Planck result for $H_0$ takes place only for sufficiently rapid ($\nu \gtrsim 100$) early transitions ($z_{\rm t} \gtrsim 5$).}
    \label{fig:H0_zdag} \vspace{7mm}
\end{figure*}

We now solve the system of~\cref{eq_phi_dless,friedmann_1_dless,friedmann_2_dless} numerically, following the method described in the Appendix. Before proceeding with the calculations, several points are worth highlighting. First, as seen in~\cref{friedmann_1_dless}, the present-day ($z=0,\, \widetilde h=1$) density parameter of the phantom field is
\begin{equation}
 \widetilde\Omega_{\phi}(z=0) \equiv\Omega_{\phi 0}=1 - \Omega_{\rm m0}-\Omega_{\rm r0}\, 
\label{phi0}   
\end{equation}
hence $\Omega_{\rm m0}$ and $\Omega_{\rm r0}$ determine $\Omega_{\phi 0}$. We assume that at the present time the potential is very flat, i.e., ${\nu(\tilde\phi_0-\tilde\phi_{\rm c})\gg1}$. Therefore, the phantom field is nearly frozen, and
$({\rm d}\tilde\phi/{\rm d}z)^2/6|_{z=0} \ll \widetilde\Omega_V(z=0)\equiv \Omega_{V 0}$. Thus, 
\begin{equation}
\Omega_{V0}\approx \Omega_{\Lambda 0}\approx \Omega_{\phi 0}\, .
\label{s0}    
\end{equation}
Subsequent numerical calculations (see the Appendix for details) confirm that these assumptions hold with very high accuracy. It is also important to note that, in this model, there are two characteristic redshift values. First, $z_{\rm t}$ marks the moment when the potential changes sign, such that $\tilde\phi(z=z_{\rm t})=\tilde\phi_{\rm c}$, leading to $V(z=z_{\rm t})=0$. It is obvious that $z_{\rm t}$ is also a potential inflection point. As we will demonstrate below, the redshift $ z_{\rm t} $ also marks key points in the evolution of the model parameters during the AdS-to-dS transition. Additionally, $ z_\dagger $ signifies the redshift at which the phantom field's energy density changes sign, satisfying $ \widetilde\Omega_{\phi}(z=z_\dagger) = 0 $. Our numerical simulations reveal that $ z_\dagger < z_{\rm t} $. This occurs because the scalar field initially rolls in the negative potential region, climbing the potential until it reaches $ V = 0 $ at $ z_{\rm t} $, and then continues ascending until the potential equals the kinetic energy $ V = X $, at which point $ \rho_\phi $ becomes zero at $ z_\dagger $.  

It is worth mentioning a few key points regarding the parameter $\nu$. One of the main goals of this article is to introduce a physical mechanism to realize $\Lambda_{\rm s}$CDM  model, which is characterized by a \textit{rapid} mirror AdS-to-dS transition in the late universe ($z\sim2$) within GR. For the type of potential given in~\cref{potential_lscdm}, it is evident that $\nu$ determines the rapidity of the transition, and a sufficiently large value of $\nu$ would result in the desired rapidity. The rapidity can be quantified by the characteristic width of the transition epoch, defined as $\Delta\tilde\phi\approx 4\nu^{-1}$, as adopted from Ref.~\cite{Akarsu:2024qsi};  the larger the $\nu$, the smaller the $\Delta\widetilde \phi$, representing the difference in the value of $\tilde\phi$ from the onset to the end of the transition. Further details on the determination of the characteristic width of the transition epoch, as well as the definitions of the onset and end of the transition, can be found in Ref.~\cite{Akarsu:2024qsi}. 

\begin{figure*}[!t]
\centering
    \begin{tabular}{@{}c@{}}\hspace{-3mm}
	\includegraphics[width=0.75\linewidth]{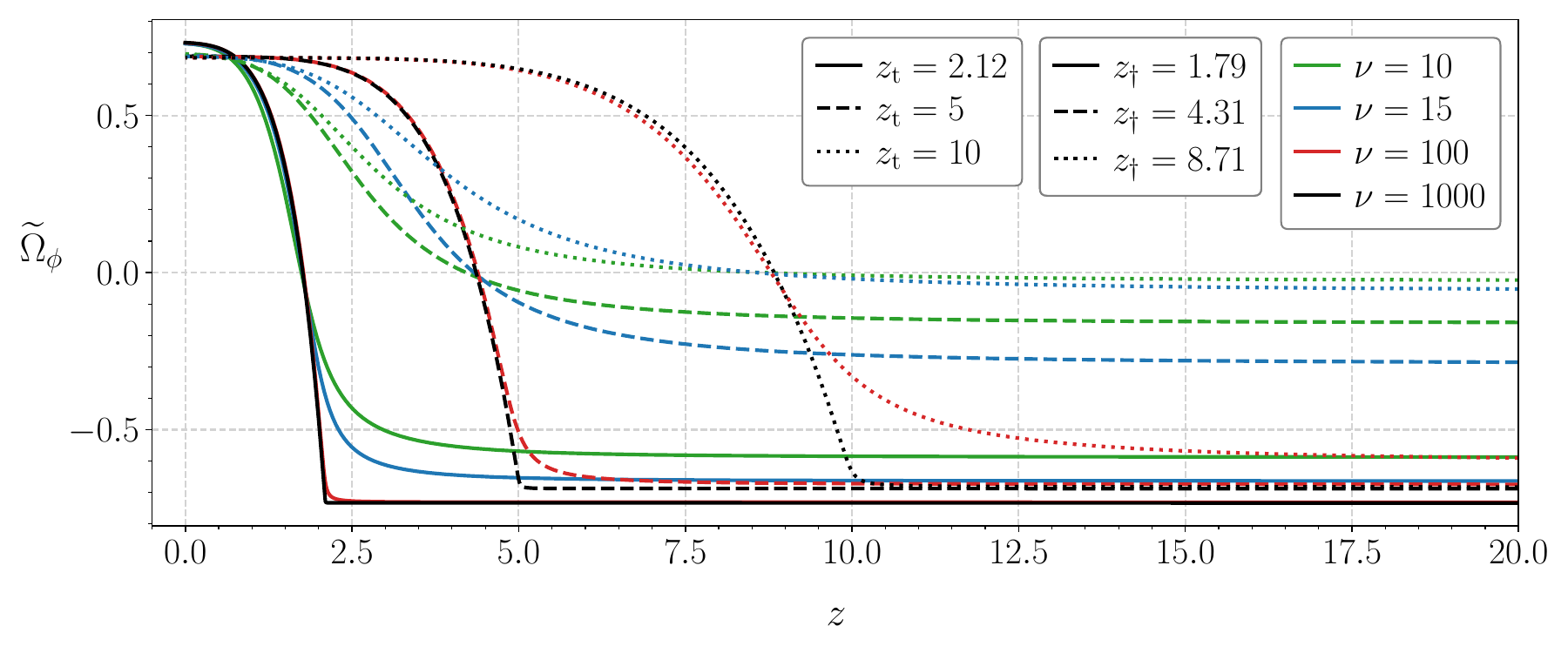}
    \end{tabular}
    \vspace{-3mm}
    \caption{Evolution of the phantom energy density  $\widetilde\Omega_{\phi}(z)=\varepsilon_{\phi}(z)/\varepsilon_{\rm{cr}0}$ for the rapidity parameters $\nu=10,15,100$ and 1000. We selected three cases where the potential changes its sign at $z_{\rm t}= 2.12, 5$ and 10. For these cases, $z_{\dagger}$ indicates the redshifts when $\widetilde\Omega_{\phi}$ crosses zero.}
\label{fig:Omega_phis} \vspace{7mm}
\end{figure*}

\begin{figure}[!t]
\centering

    \begin{tabular}{@{}c@{}}
	\includegraphics[width=1\linewidth]{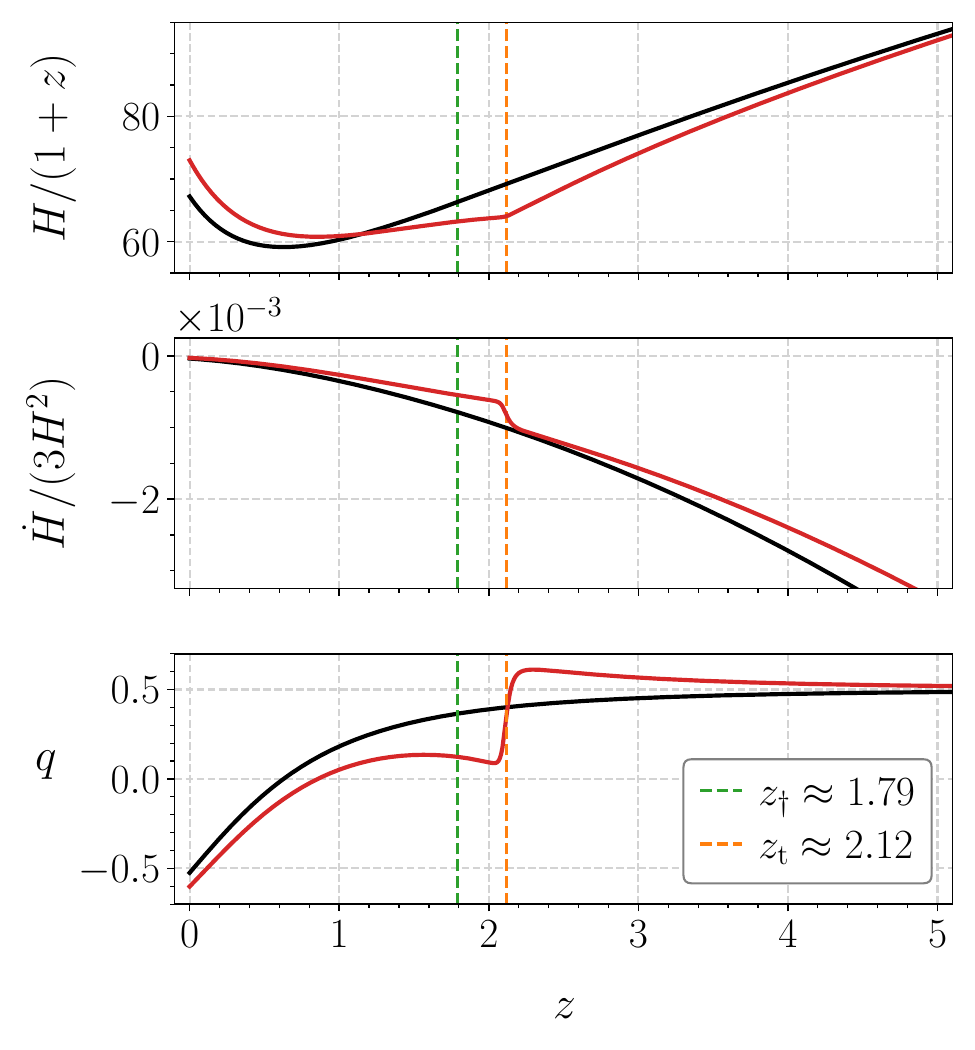}
    \end{tabular}

    \vspace{-3mm}

    \caption{The Hubble parameter $H/(1+z)~{\rm km\, s^{-1}\, Mpc^{-1}}$ , the dimensionless combination $\dot H/(3H^2)$, where $\dot H=dH/dt$,  and the deceleration parameter $q$ as functions of the redshift $z$. 
    $z_{\rm t}$ determines the moment of the transition where the potential is equal to zero: $V(z=z_{\rm t})=0$ and $z_\dagger$ corresponds to the redshift at which the phantom energy density is equal to zero: $\widetilde\Omega_{\phi}(z=z_\dagger)=0$. Black curves describe the standard $\Lambda$CDM model. }
\label{fig:hubbles_v_100}
\end{figure}

\begin{figure*}[!ht]
\centering

    \begin{tabular}{@{}c@{}}\hspace{-3mm}
	\includegraphics[width=.49\linewidth]{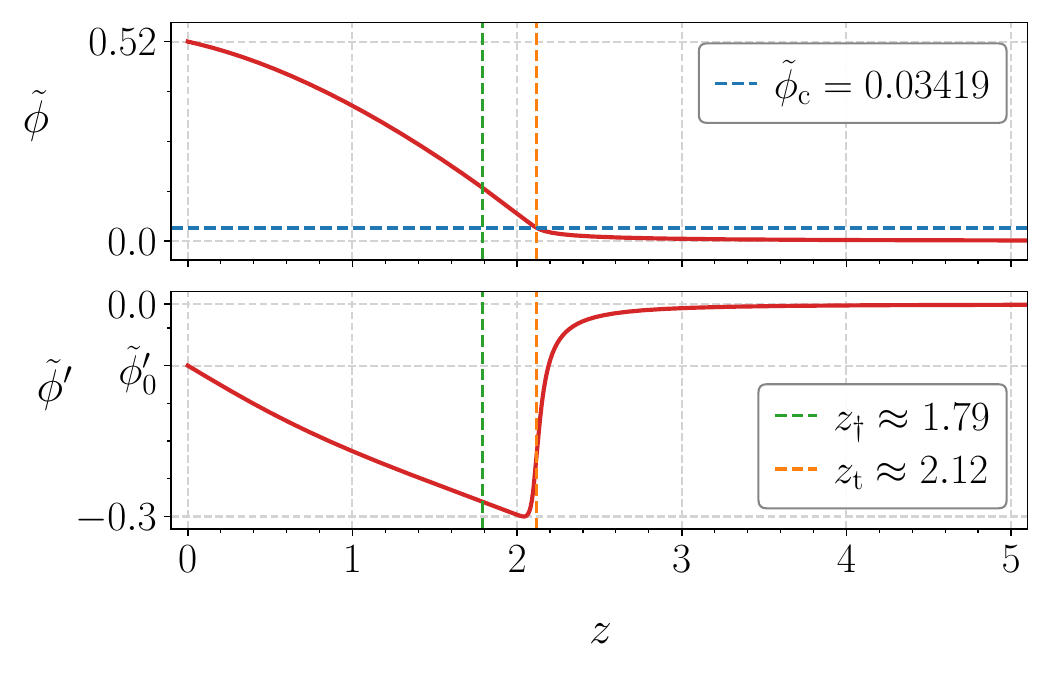}
    \end{tabular}
    \begin{tabular}{@{}c@{}}\hspace{2mm}
	\includegraphics[width=.49\linewidth]{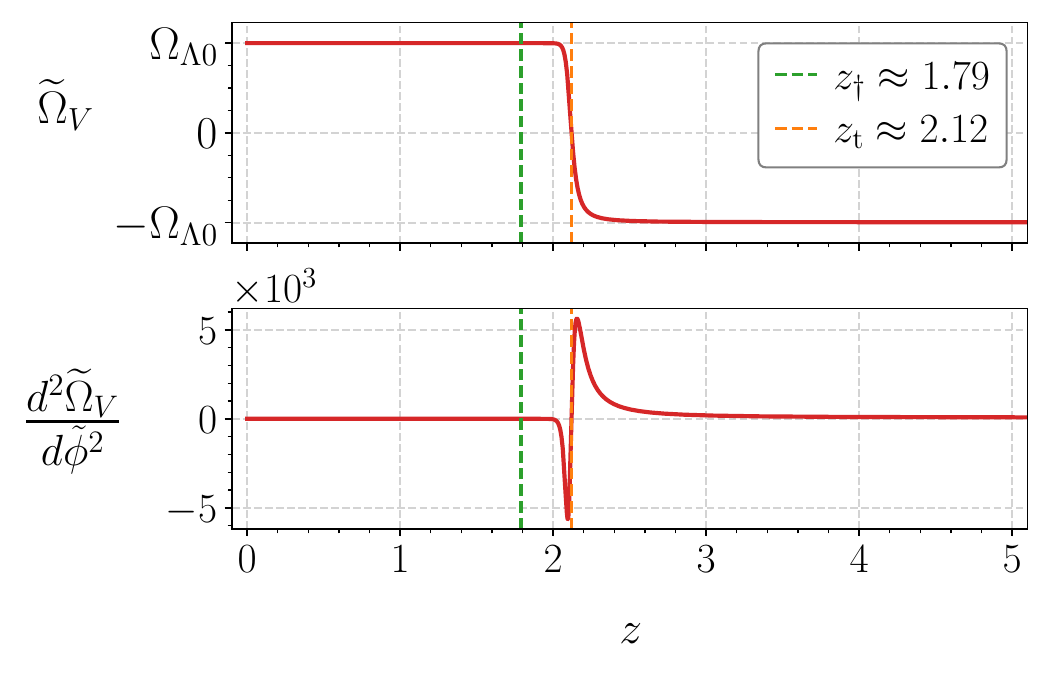}
    \end{tabular}

    \vspace{-3mm}

    \caption{On the left panel: The dimensionless phantom field $\tilde\phi$ and its derivative $\tilde\phi'\equiv d\tilde\phi/dz$. The value of this derivative at present time $\tilde\phi'_0 = -0.09661$. $\tilde\phi_{\rm c}=0.03419$ is the value of the phantom field where the potential changes sign. On the right panel: The dimensionless potential $\widetilde\Omega_V = V/\varepsilon_{\rm cr0}$ and its second derivative where $\Omega_{\Lambda 0}=0.73161$.}
\label{fig:phi_and_potential_1} \vspace{7mm}
\end{figure*}

\begin{figure*}[!ht]
\centering
    \begin{tabular}{@{}c@{}}\hspace{-3mm}
	    \includegraphics[width=.49\linewidth]{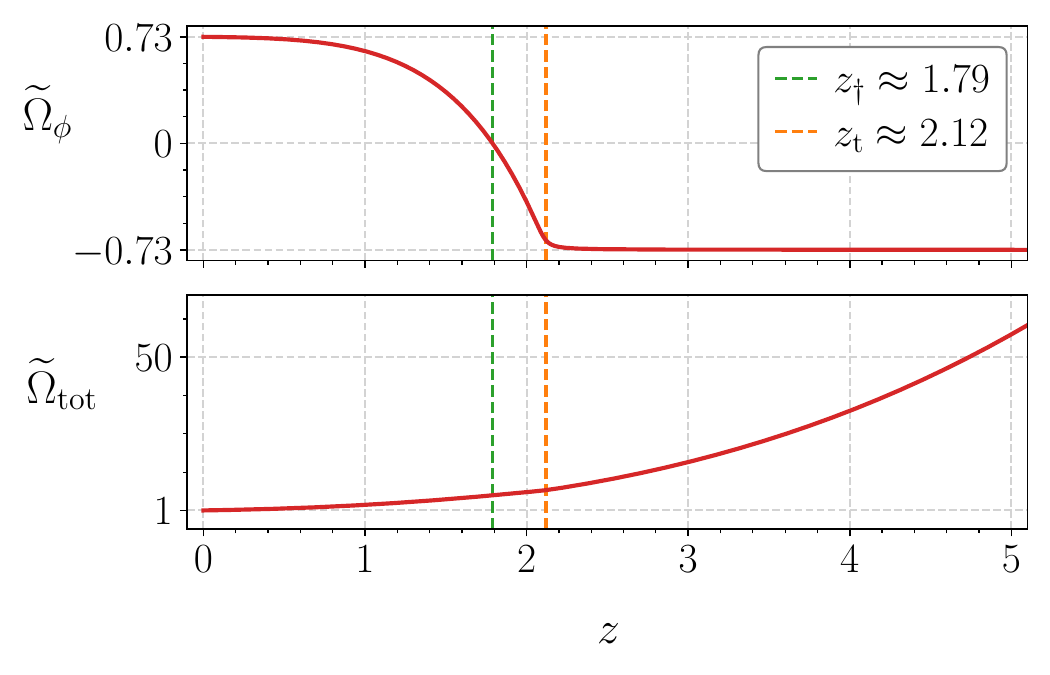}

    \end{tabular}
    \begin{tabular}{@{}c@{}}\hspace{2mm}
	\includegraphics[width=.49\linewidth]{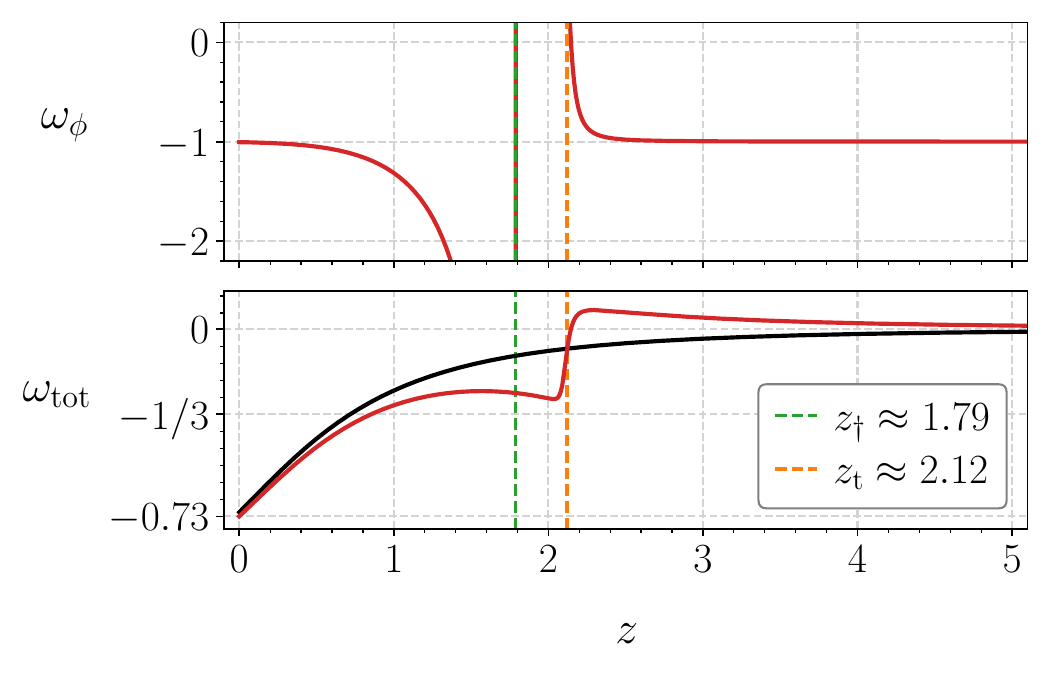}
    \end{tabular}

    \vspace{-3mm}

    \caption{On the left panel: the EoS parameter $\omega_{\phi}$ for the phantom field and the EoS parameter $\omega_{\rm tot}$ for total matter (phantom + CDM + radiation). On the right panel: the dimensionless energy density $\widetilde\Omega_{\phi}=\varepsilon_{\phi}/\varepsilon_{\rm cr0}$ for the phantom and the dimensionless energy density $\widetilde\Omega_{\rm tot}=\varepsilon_{\rm tot}/\varepsilon_{\rm cr0}$ for total matter. The total energy density is everywhere positive.}
\label{fig:omegas_Omegas_1}
\end{figure*}

The result of the numerical integration of the system of~\cref{eq_phi_dless,friedmann_1_dless,friedmann_2_dless} for the phantom field (by setting $\xi=-1$ and parameter $\xi_1=-1$) is presented in~\cref{tab1} and~\cref{fig:H0_zdag,fig:Omega_phis,fig:hubbles_v_100,fig:phi_and_potential_1,fig:omegas_Omegas_1}. The initial conditions are $\tilde\phi_{\rm{in}} =\tilde\phi'_{\rm{in}} =0$ (see the Appendix for details). We observe a consistent delay in the sign change of the energy density of the phantom field relative to the sign change of the potential, such that $z_\dagger < z_{\rm t}$, with these two parameters being highly correlated---a Pearson correlation coefficient of approximately $r\sim0.99$, demonstrating a nearly perfect linear correlation between these two parameters. This behavior is expected, as the sign change in the potential drives the sign change in the energy density. Furthermore,~\cref{tab1} reveals a negative correlation between $H_0$ and both $z_{\rm t}$ and $z_\dagger$---shown for some selected values of $\nu$. In other words, smaller values of $z_{\rm t}$ or $z_\dagger$ correspond to larger values of $H_0$, as anticipated. This is explained as follows: Since the pre-recombination universe in our model remains consistent with the $\Lambda$CDM framework, the comoving sound horizon at last scattering, $r_* = \int_{z_*}^{\infty} c_{\mathrm{s}} H(z)^{-1} \mathrm{d}z$, where $z_* \sim 1090$ is the redshift of the last scattering surface and $c_{\mathrm{s}}$ is the sound speed in the photon-baryon plasma, remains essentially unchanged from its $\Lambda$CDM counterpart. From the Planck CMB spectra, the angular scale of the sound horizon, $\theta_* = r_*/D_M(z_*)$, and the present-day physical matter density, $\Omega_{\rm m0} h^2$ (derived from the peak structure and damping tail, where $h \equiv H_0/100~\mathrm{km\,s^{-1}\,Mpc^{-1}}$), are both measured, almost model-independently, with high precision. Consequently, in our model, namely, a phantom field driven $\Lambda_{\rm s}$CDM scenario, both the comoving angular diameter distance to last scattering, $D_{M}(z_*) = c \int_{0}^{z_*} H(z)^{-1} \mathrm{d}z$, and $\Omega_{\rm m0} h^2$ are expected to align with their Planck-inferred values for $\Lambda$CDM. Therefore, any suppression of $H(z)$ (i.e., $H(z) < H(z)_{\Lambda\rm CDM}$) for $z > z_\dagger$, due to the negative phantom energy density in this regime, must be offset by an enhancement ($H(z) > H(z)_{\Lambda\rm CDM}$) at lower redshifts, $z < z_\dagger$, to maintain consistency with the $\Lambda$CDM inferred comoving angular diameter distance. This implies an increased $H_0\equiv H(z=0)$ and thereby also a decreased $\Omega_{\rm m0}$ compared to the Planck-$\Lambda$CDM model (i.e, standard $\Lambda$CDM model limited only by Planck CMB data~\cite{Planck:2018vyg}). Consequently, a later transition---i.e., smaller values of $z_\dagger$ (and correspondingly $z_{\rm t}$)---results in a longer and more pronounced period after recombination during which $H(z) < H(z)_{\Lambda\rm CDM}$. As result, decreasing $z_\dagger$  (or $z_{\rm t}$) leads to a greater enhancement of $H_0$ and a corresponding reduction in $\Omega_{\rm m0}$ compared to the predictions of the Planck-$\Lambda$CDM model, provided that the transition occurs before the negative energy density of the phantom field completely dominates the universe---an event that would halt expansion and trigger contraction.

In~\cref{fig:H0_zdag}, we investigate the dependence of the Hubble constant on $z_{\rm t}$ and $z_\dagger$ for different values of the rapidity parameter $\nu$. The analysis shows that, using some selected values of $\nu$, the Hubble constant attains the SH0ES measurement of $H_0 = 73.04 \pm 1.04~{\rm km\, s^{-1}\, Mpc^{-1}}$~\cite{Riess:2021jrx} (see also~\cite{Uddin:2023iob,Breuval:2024lsv}) for $z_{\rm t} \approx 2.1$ (correspondingly for $z_{\dagger} \sim 1.8$ ), irrespective of the rapidity of the transition. As $z_{\rm t}$ (or $z_\dagger$) increases, the predicted value of $H_0$ decreases and asymptotically converges to a specific minimum value, denoted as $H_0^{\rm min}$ ($H_0^{\rm min}=\lim_{z_{\rm t}\to\infty} H_0$). This minimum value, $H_0^{\rm min}$, is found to depend on the rapidity of the transition, determined by the parameter $\nu$. For fast transitions ($\nu = 100, 1000$), $H_0^{\rm min}$ aligns with the Planck-$\Lambda$CDM prediction of $H_0 = 67.4 \pm 0.5~{\rm km\, s^{-1}\, Mpc^{-1}}$~\cite{Planck:2018vyg}. In contrast, for slow transitions ($\nu = 10, 15$), $H_0^{\rm min}$ remains above the Planck-$\Lambda$CDM prediction. Specifically, for $\nu = 15$, $H_0^{\rm min}$ reaches $68.0~{\rm km\, s^{-1}\, Mpc^{-1}}$, matching the upper limit of the Planck-$\Lambda$CDM prediction, while for $\nu = 10$, it increases further to $68.5~{\rm km\, s^{-1}\, Mpc^{-1}}$. In what follows, we will explain the relationships observed between $H_0$ and the parameters $z_\dagger$ and $z_{\rm t}$.

It is well established that conventional phantom models (say, {\it p}-phantom) can provide only partial alleviation of the \( H_0 \) tension; see Refs.~\cite{Vagnozzi:2018jhn,DiValentino:2019exe,DiValentino:2019dzu,Alestas:2020mvb,Vazquez:2020ani,Banerjee:2020xcn,Heisenberg:2022gqk,Lee:2022cyh}. In conventional phantom models, the energy density decreases with redshift while remaining positive, leading to a moderate suppression of $H(z)$ at high redshifts and a correspondingly modest increase at lower redshifts, including in $H_0$. Similarly, our phantom model—characterized by a late-time transition from {\it n}-phantom to {\it p}-phantom—also exhibits a decreasing energy density with redshift. However, unlike conventional phantom models, it has a distinguishing feature: as we go into the past, the energy density crosses below zero at a certain redshift, $z = z_\dagger$, and remains negative thereafter. This negative phase amplifies the suppression of $H(z)$ at earlier times, making the subsequent compensatory increase in $H(z)$ at lower redshifts more pronounced compared to conventional phantom models. Consequently, our model leads to a significantly greater increase in $H_0$ than standard phantom scenarios, even though the phantom field behaves like a positive CC at low redshifts. These similar yet distinct features between our model and conventional phantom models also explain why, in our model, slow transitions occurring at large redshifts, viz., $z_{\dagger}\gtrsim 4$ (or $z_{\rm t}\gtrsim 4.5$), predict $H_0$ values that remain above the Planck-$\Lambda$CDM prediction, whereas fast transitions at large redshifts yield $H_0$ values consistent with Planck-$\Lambda$CDM. In the case of a fast transition, e.g., $\nu = 100,\,1000$, if the transition occurs too early, e.g., $z_{\dagger}\sim4,\,9$ (corresponding to $z_{\rm t}\sim5,\,10$, see~\cref{fig:Omega_phis})---that is, when pressureless matter is still dominant---the phantom field in its negative energy density phase lacks sufficient time and strength to significantly influence $H(z)$; consequently, $H(z)$ remains largely consistent with $\Lambda$CDM during that period. Later, when the phantom field begins to dominate, it already closely resembles a CC---namely, as shown in~\cref{fig:Omega_phis}, $\widetilde\Omega_\phi\approx 0.7$ for $z\lesssim2.5$ when $z_{\rm t}=5$ (with $z_\dagger=4.31$) and for $z\lesssim5.0$ when $z_{\rm t}=10$ (with $z_\dagger=8.71$). As a result, the overall impact of the phantom field on $H(z)$ is minimal, and the model effectively resembles $\Lambda$CDM, thereby predicting an $H_0$ value consistent with Planck-$\Lambda$CDM. Now consider the case of a slow transition (e.g., $\nu = 10,\,15$). If the transition occurs too early, e.g., considering again $z_{\dagger}\sim4,\,9$ (corresponding to $z_{\rm t}\sim5,\,10$), the negative energy density phase of the phantom field again has minimal influence on $H(z)$ during the matter-dominated era, similar to the case of an early fast transition. However, in this scenario, due to the slower nature of the transition, after entering the positive energy density regime, the phantom field does not settle into a horizontal plateau but instead continues to increase significantly, even up to the present time.  As shown in~\cref{fig:Omega_phis}, for $\nu = 10,15$, the phantom energy density $\widetilde\Omega_\phi$ remains positive for $z<z_\dagger=4.31$ choosing $z_{\rm t}=5$ and for $z<z_\dagger=8.71$ choosing $z_{\rm t}=10$. However, unlike a CC, $\widetilde\Omega_\phi$ does not remain constant at low redshifts but instead continues to increase at a significant rate as redshift decreases---a behavior characteristic of conventional phantom dark energy models.  In other words, when the phantom field begins to influence $H(z)$, its EoS parameter remains significantly below $-1$, meaning that overall, our model effectively resembles conventional phantom models with positive energy densities rather than $\Lambda$CDM, and hence $H_0^{\rm min}$ remains above the Planck-$\Lambda$CDM value.

 So far, we have elaborated on the model predictions for increasing values of $z_{\rm t}$ (or $z_\dagger$) beyond $z_{\rm t} \sim 2.1$ (or $z_\dagger \sim 1.8$), where our model predicts $H_0$ values consistent with the SH0ES measurements~\cite{Riess:2021jrx,Uddin:2023iob,Breuval:2024lsv}, regardless of whether the transition is slow or fast. We now consider the consequences of decreasing $z_{\rm t}$ (or $z_\dagger$). To this end, assume that, due to transition delays, the scalar field remains in its flat AdS potential regime for an extended period, so that the total energy density may be approximated by $\widetilde{\Omega}_{\rm tot}(z) \approx \widetilde{\Omega}_{\rm m} + \widetilde{\Omega}_V \approx \Omega_{\rm m0}(1+z)^3 - \Omega_{\Lambda0} = \Omega_{\rm m0}(1+z)^3 - (1-\Omega_{\rm m0})$. Consequently, $\widetilde{\Omega}_{\rm tot}(z)$ becomes zero at 
 \begin{eqnarray}
 z_{\rm s} = \bigl(\Omega_{\rm m0}^{-1}-1\bigr)^{1/3} - 1\,,    
 \end{eqnarray}
  which yields $z_{\rm s} \sim 0.33$ for $\Omega_{\rm m0} \sim 0.3$. Thus, to ensure an ever-expanding universe, we must have $z_\dagger \gtrsim 0.33$. However, focusing on realistic cases,~\cref{tab1} and~\cref{fig:H0_zdag} show that $H_0$ becomes very sensitive to $z_{\rm t}$ and increases rapidly---exceeding approximately $75~{\rm km\,s^{-1}\,Mpc^{-1}}$---when $z_{\rm t} \lesssim 2$ (corresponding to $z_\dagger \lesssim 1.6$). For example, for $\Omega_{\rm m0}=0.25$ the model yields $H_0 \approx 75~{\rm km\,s^{-1}\,Mpc^{-1}}$ for $z_{\rm t}\approx 1.9$, corresponding to $z_\dagger \approx 1.5$–$1.6$. Therefore, for a realistic cosmological model---where $H_0$ does not exceed the largest local measurement values (roughly $75~{\rm km\,s^{-1}\,Mpc^{-1}}$)---we would expect $z_{\rm t}\gtrsim 1.8$ and, correspondingly, $z_\dagger \gtrsim 1.5$. Our estimates for \(z_\dagger\) in the present phantom field realization of \(\Lambda_{\rm s}\)CDM are in good agreement with previous results obtained in the gDE\cite{Akarsu:2019hmw} and \(\Lambda_{\rm s}\)CDM models~\cite{Akarsu:2021fol,Akarsu:2022typ,Akarsu:2023mfb,Akarsu:2024eoo,Yadav:2024duq}.

Interestingly, these lower bounds on $z_{\dagger}$ and $z_{\rm t}$ closely match those obtained when requiring that the time derivative of the Hubble parameter, $\dot{H}$, remains negative for all redshifts. In fact, from Eq.~\eqref{H dot dless} we have $\dot{H} = -\frac{1}{2} H_0^2 \bigl( 3\,\widetilde{\Omega}_{\rm m} + 4\,\widetilde{\Omega}_{\rm r} + 6\,\widetilde{\Omega}_K \bigr)$, where the matter and radiation contributions satisfy $\widetilde{\Omega}_{\rm m},\,\widetilde{\Omega}_{\rm r} > 0$ while the kinetic contribution of the phantom field satisfies $\widetilde{\Omega}_K < 0$. Since the maximum (in absolute value) kinetic energy density of the phantom field occurs during the transition, if $\dot{H}$ is negative at $z \sim z_{\rm t}$, it will remain negative at all other redshifts. During the transition, the potential given by Eq.~\eqref{potential_lscdm} is well approximated by a linear function,
\begin{equation}
V(\phi) \approx \Lambda\,\nu\,\sqrt{\kappa}\,(\phi-\phi_c)\,,   
\end{equation}
so that the Klein--Gordon equation~\eqref{fielddyneq} simplifies to
\begin{equation}
\ddot{\phi} \approx -3H\,\dot{\phi} + c^2 \Lambda\,\nu\,\sqrt{\kappa}\,.    
\end{equation}
 This equation describes the evolution of $\phi$ under the influence of the cosmic ``friction'' ($-3H\,\dot{\phi}$) and a constant external ``force'', which we define as $A \equiv c^2 \Lambda\,\nu\,\sqrt{\kappa}$. Let $\Delta\tau$ denotes the (short) duration of the transition. Since $H(z_{\rm t})\,\Delta\tau \ll 1$ (with $H(z_{\rm t})$ being approximately the inverse of the age of the universe at the transition epoch), the friction term is initially subdominant, and we may approximate $\dot{\phi} \approx A\,t$, assuming the pre-transition value of $\dot{\phi}$ is negligible. Taking the dimensionless field variation during the transition as $\Delta\widetilde{\phi} \approx 4/\nu$ (see, e.g., Ref.~\cite{Akarsu:2024qsi}), one finds $\Delta\tau \approx \bigl[2\Delta\widetilde{\phi}/(A\sqrt{\kappa})\bigr]^{1/2} \approx \bigl[8/(A\sqrt{\kappa}\,\nu)\bigr]^{1/2}$. Thus, the maximum field derivative at the end of the transition is $\dot{\phi}_{\rm max} \approx A\,\Delta\tau \approx 2\sqrt{2\,c^2\Lambda}$ and the corresponding (rescaled) kinetic energy density is estimated as $-\widetilde{\Omega}_K = (\dot{\phi}_{\rm max})^2/(2c^2\varepsilon_{\rm cr0}) \approx 4\,\Omega_{\Lambda0}$. At the end of the transition, the Friedmann--Raychaudhuri equation then yields $\dot{H}(z_{\rm t}) \approx -(3H_0^2/2)\bigl[ \widetilde{\Omega}_{\rm m}(z_{\rm t}) - 8\,\Omega_{\Lambda0}\bigr]$. Since the matter density scales as $\widetilde{\Omega}_{\rm m}(z_{\rm t}) = \Omega_{\rm m0}\,(1+z_{\rm t})^3$ and, neglecting radiation, the flat-universe condition implies $\Omega_{\rm m0} + \Omega_{\Lambda0} = 1$ (i.e. $\Omega_{\Lambda0} = 1-\Omega_{\rm m0}$), the requirement for $\dot{H}(z \sim z_{\rm t}) < 0$ translates into $\Omega_{\rm m0}\,(1+z_{\rm t})^3 > 8\,(1-\Omega_{\rm m0})$, or equivalently,
\begin{equation}
    z_{\rm t} \gtrsim 2\Bigl(\Omega_{\rm m0}^{-1}-1\Bigr)^{\frac{1}{3}}-1 \,.
\label{z-lim} 
\end{equation}
For example, taking $\Omega_{\rm m0} \sim 0.27$ yields $z_{\rm t} \gtrsim 1.8$. In our model, a combination such as $\Omega_{\rm m0}\sim0.27$ with $z_{\rm t}\sim2.1$ (the value required for $H_0$ to match the SH0ES $H_0$ measurement, as shown in~\cref{tab1} and~\cref{fig:H0_zdag}) ensures that $\dot{H}$ remains negative at all redshifts (as seen also in~\cref{fig:hubbles_v_100}). However, if $z_{\rm t}$ approaches the lower limit determined by~\cref{z-lim}, the value of $\dot{H}(z_{\rm t})$ approaches zero and may even turn positive. As seen from in~\cref{tab1} and~\cref{fig:H0_zdag}, it may occur when $H_0$ becomes highly sensitive to $z_{\rm t}$ and increases rapidly, exceeding approximately 75~km\,s$^{-1}$Mpc$^{-1}$. For instance, for $\Omega_{\rm m0}=0.25$, the model yields $H_0\approx75$~km\,s$^{-1}$Mpc$^{-1}$ for $z_{\rm t}\approx1.9$, while our condition requires $z_{\rm t} \gtrsim 1.88$. It is particularly interesting---if not coincidental---that the largest local $H_0$ measurements, which are close to the upper bound of the SH0ES result, correspond to the threshold at which $\dot{H}$ can go from negative to positive in our model. In other words, if one requires that $\dot H$ to remain negative at all redshifts, our model predict a maximum value of $H_0\approx75$~km\,s$^{-1}$Mpc$^{-1}$.

Thus, we have discussed in detail how the smooth transition of the phantom field’s potential from an AdS-like to a dS-like form—resulting in a sign change in its energy density around $z_{\dagger}$—affects the prediction of $H_0$ depending on when and how rapidly these transitions occur. To conclude this section and to highlight the characteristic features of our model as well as its similarities and deviations from the standard $\Lambda$CDM model, we illustrate the expansion dynamics of the universe and the scalar field, as well as the total physical ingredients of the universe, for a fast transition (i.e., $\nu=100$) in~\cref{fig:hubbles_v_100,fig:phi_and_potential_1,fig:omegas_Omegas_1}. We ensure consistency with the CMB power spectra~\cite{Planck:2018vyg} and the SH0ES $H_0$ measurement of $73.04~\mathrm{km\,s^{-1}\,Mpc^{-1}}$~\cite{Riess:2021jrx}; our numerical analysis yields $z_{ \rm t}\approx 2.12$ and $z_{\dagger }\approx 1.79$ (using the initial conditions $\tilde\phi_{\rm in}=\tilde\phi'_{\rm in}=0$ for the scalar field).

\cref{fig:hubbles_v_100} demonstrates the typical behavior of the Hubble parameter, its time derivative, and the deceleration parameter for our model. The black curves represent the standard $\Lambda$CDM model, and all curves are smooth. Notably, we observe that $H(z)$ is suppressed for $z \gtrsim 1.7$ relative to $\Lambda$CDM, with a compensating enhancement at lower redshifts. The deceleration parameter $q$ rises noticeably above the value $q=\frac{1}{2}$ (characteristic of a matter-dominated universe) when the negative dark energy of the phantom field becomes significant compared to pressureless matter—as $z$ increases and approaches $z_{\rm t}$. During the transition, $q(z)$ rapidly decreases to values close to zero and remains near zero throughout the approximately linear expansion phase in the range $z\sim2$ to $z\sim1$, whereas the $\Lambda$CDM model continues to experience decelerated expansion during this period. Finally, the universe enters the accelerated expansion phase slightly earlier than in the $\Lambda$CDM model. At late times (i.e., $z\lesssim1$), our scalar field has settled into the dS-like potential regime, (namely, the flat, positive potential regime), effectively mimicking a positive CC with $\omega_\phi\sim-1$ and $\widetilde\Omega_\phi\sim0.7$ (see~\cref{fig:omegas_Omegas_1}). Consequently, for $z\lesssim1$ our model behaves essentially like $\Lambda$CDM, albeit with a higher $H(z)$, which leads to a larger $H_0$ and a smaller $q(z)$. We also observe that $\dot{H}$ remains negative at all redshifts.

 In the left panel of~\cref{fig:phi_and_potential_1}, we show the evolution of the phantom scalar field in dimensionless form ($\tilde\phi \equiv \sqrt{\kappa}\phi$), specifically its position $\tilde\phi$ and its derivative with respect to redshift, $\tilde\phi' = \frac{d\tilde\phi}{dz}$. In the right panel, we display the dimensionless potential, $\widetilde\Omega_V = V/\varepsilon_{\rm cr0}$ (i.e., the potential scaled by the present-day critical energy density), along with its curvature, $\frac{d^2\widetilde\Omega_V}{d\tilde\phi^2}$, as functions of $z$. At early times, when the field is in the AdS-like flat potential regime, it is in a slow-rolling phase ($\tilde\phi' \approx 0$). As the field enters the rapidly increasing region of the potential, its speed quickly rises (with $|\tilde\phi'|$ reaching approximately 0.3), signaling the onset of the transition epoch. Upon passing the inflection point of the potential at $z = z_{\rm t}$, the field attains its maximum speed and then begins to decelerate. Finally, when the field reaches the dS-like positive flat potential, its speed continues to decrease as the potential increasingly dominates over the kinetic term, and the field closely mimics a positive CC for $z \lesssim 1$. The right panel of~\cref{fig:phi_and_potential_1} demonstrates that the potential $\widetilde\Omega_V$ tends asymptotically to $\Omega_{\Lambda0}$ when $z\to 0$. For the chosen parameters, the calculated value is $\Omega_{\Lambda0}=0.73161$; and, according to~\cref{eq:Omega_M0}, the corresponding cosmological parameter is $\Omega_{\rm m0}=0.26831$, so that the relation $ \Omega_{\Lambda0}+\Omega_{\rm m0}\approx1$ holds with high accuracy. Notably, the curvature of the potential is nearly zero before the onset of the AdS–dS transition; however, at the start of the transition it rapidly increases to a finite positive value, then, after the field reaches transition at $z_{\rm t}$, the curvature quickly vanishes and becomes negative before approaching zero again once the field settles into the dS-like flat potential. Given that the potential’s curvature is nearly zero both before and after the transition—and only becomes significantly positive for a brief period ($\Delta z \lesssim 0.5$) at the onset---the effective mass of some modes of perturbations may momentarily become imaginary, but it quickly returns to a real value. Due to this transient behavior our model can remain perturbatively stable\footnote{In our model, the effective mass squared for perturbations~\cite{Carroll:2003st} is given by $m_{\rm eff}^2(z) \sim k^2 - a^2 \frac{d^2V}{d\phi^2}  = k^2 - \frac{3}{(1+z)^2} \frac{H_0^2}{c^2} \frac{d^2\widetilde{\Omega}_V}{d\widetilde\phi^2}$, where we set $ a_0 = 1 $, and $ k $ is the comoving wavenumber. For the SH0ES $ H_0 $ measurement and $ z_{\rm t} = 2.12 $, the prefactor in front of the second derivative is $ \sim2\times 10^{-47} \, {\rm cm}^{-2} $. According to our calculations, the maximum value of this derivative is of the order of $ 5\times 10^{3} $ (see~\cref{fig:phi_and_potential_1}). Therefore, for modes satisfying $ k^2 \lesssim 10^{-43} \, {\rm cm}^{-2} $, the effective mass squared can become negative. However, this occurs only for a brief period, meaning that perturbations may not have sufficient time to develop significantly. In this paper, we analyze the phantom scalar field model within the $\Lambda_{\rm s}$CDM paradigm at the background level. A full perturbative analysis will be developed in a separate work.}.
 
In the left panel of~\cref{fig:omegas_Omegas_1}, we present the dimensionless energy density parameter of the phantom field, $\widetilde\Omega_{\phi} = \varepsilon_{\phi}/\varepsilon_{\rm cr0}$, along with the total dimensionless energy density parameter, $\widetilde\Omega_{\rm tot} = (\varepsilon_{\phi} + \varepsilon_{\rm m} + \varepsilon_{\rm r})/\varepsilon_{\rm cr0}$, which accounts for the combined contributions of the phantom field, pressureless matter (CDM and baryons), and radiation. The right panel focuses on the EoS parameters, $\omega_\phi$ and $\omega_{\rm tot}$, which correspond to the phantom field and the total energy content of the universe, respectively.  
At high redshifts, specifically for $z\gtrsim z_{\rm t}=2.12$, $\widetilde\Omega_{\phi} \approx -\widetilde\Omega_{\phi0} \approx -0.73$ with $\omega_\phi \approx -1$, indicating that the scalar field is in a slow-roll phase (i.e., the potential dominates over the kinetic term) within the AdS-like flat potential. Consequently, the scalar field effectively behaves as a negative CC during this epoch. Once the transition epoch begins, $\widetilde\Omega_{\phi}$ starts to increase, approaching zero, while its EoS parameter rises rapidly. As $\widetilde\Omega_{\phi}$ smoothly transitions from negative to positive values, the EoS parameter exhibits a singularity—diverging to $+\infty$ before reemerging from $-\infty$—after which it rapidly increases, asymptotically approaching $-1$. This is a 'safe' singularity because the divergence arises purely from the definition of the EoS parameter—specifically, the energy density in the denominator momentarily passes through zero while the pressure in the numerator remains finite—ensuring that no fundamental physical quantity actually diverges. Furthermore, as demonstrated in~\cref{sec:model}, despite the unusual behavior of the phantom field’s EoS parameter, the speed of sound associated with the phantom field always remains equal to the speed of light. As a result, this singularity does not pose any threat to the consistency of the FRW universe.  In fact, $\widetilde\Omega_{\rm tot}$, which also includes contributions from pressureless matter, always remains positive and decreases smoothly as $z$ decreases, asymptotically approaching zero (as shown in the lower-left panel of~\cref{fig:omegas_Omegas_1}). Consequently, $\omega_{\rm tot}$ never exhibits a singularity. Moreover, as shown in~\cref{fig:omegas_Omegas_1}, the total EoS parameter, $\omega_{\rm tot} = p_{\rm tot}/\varepsilon_{\rm tot}$, remains greater than $-1$, and the total energy density parameter, $\widetilde\Omega_{\rm tot}$, stays positive for all redshifts. This ensures that the condition $p_{\rm tot} + \varepsilon_{\rm tot} > 0$ is satisfied alongside $\varepsilon_{\rm tot} > 0$, thereby guaranteeing the fulfillment of the weak energy condition (WEC) throughout cosmic evolution. More precisely, our model satisfies the strong form of the weak energy condition (S-WEC), reinforcing its physical viability.

It is well known that in GR, the total EoS parameter, which describes the overall content of the universe, and the deceleration parameter are related by $q(z) = \frac{1}{2}+\frac{3}{2}\omega_{\rm tot}(z)$ in a spatially flat FRW universe. Therefore, the behavior of $\omega_{\rm tot}(z)$  parallels exactly that of $q(z)$.  We observe that $\omega_{\rm tot} \approx 0$ at high redshifts ($z \gtrsim 4$), as expected in the matter-dominated era. It then rises noticeably, particularly when the negative energy of the phantom field becomes significant, until the onset of the transition at $z_{\rm t} = 2.12$. Once the AdS–dS transition epoch begins, $\omega_{\rm tot}$ drops rapidly to around $-1/3$, where it remains until $z \sim 1$. Eventually, it decreases further, reaching $\sim -0.73$ today ($z = 0$), a value very close to that of the $\Lambda$CDM model. Thus, the total EoS parameter remains smooth and free of singularities, confirming that the pressure of the phantom field does not diverge. The singularity in the phantom field’s EoS arises solely due to its energy density passing through zero, while the total energy density remains strictly positive at all times. Finally, for $z \lesssim 1$, we find $\widetilde\Omega_{\phi} \approx 0.73$ and $\omega_\phi \approx -1$, which reaffirms that our model is essentially indistinguishable from $\Lambda$CDM at late times, even though it predicts an $H_0$ value consistent with the SH0ES measurements~\cite{Riess:2021jrx,Uddin:2023iob,Breuval:2024lsv}.

\section{General transition models: Dynamics and Cosmological Effects}
\label{sec:General_transition_models}

In this section, we consider the general transition model described by the potential in~\cref{eqn:gen_pot}, which asymptotically approaches a dS-like positive flat potential in the late universe—effectively mimicking a positive CC today. However, unlike the specific model analyzed earlier, this general formulation allows for an initial flat potential with an arbitrary value, which can be either negative or positive and may be larger or smaller than its present-day value. Thus, we treat $\xi_1$ as a free parameter while also allowing for different choices of $\xi = \pm 1$ to account for both phantom and quintessence scenarios. In this framework, the transitions described by~\cref{eqn:gen_pot} (or its dimensionless counterpart in~\cref{parameters}) can occur for both phantom and quintessence fields. For a quintessence field, we require $\xi_1 > 1$, leading to a transition from a higher to a lower field potential (i.e., from $\xi_1 \Lambda$ to $\Lambda$ for $\Lambda > 0$). 

We analyze the following cases for both quintessence ($\xi = 1$) and phantom ($\xi = -1$) models:
\begin{enumerate}
    \item \textbf{General AdS-to-dS transitions} ($\xi_1 < 0$): In this class of models, we focus on the phantom field case ($\xi = -1$), as such a transition—where the field evolves from the bottom to the top of its potential—is only possible for phantom fields. For definiteness, we compare two representative cases: $\xi_1 = -1$ (mirror AdS-to-dS transition) and $\xi_1 = -0.5$. 

    \item \textbf{0-dS transition} ($\xi_1 = 0$): This scenario, where the potential transitions from an extremely small initial value (ideally zero as $\tilde\phi \to -\infty$) to a positive value $\Lambda$, corresponds to the emergence of a cosmological constant. This transition is only possible for the phantom field ($\xi = -1$).

    \item \textbf{General dS-to-dS transitions} ($\xi_1 > 0$): In this category, we distinguish between two cases: 
    \begin{itemize}
        \item Phantom models ($\xi = -1$), where the scalar field climbs up its potential ($\xi_1 < 1$).
        \item Quintessence models ($\xi = 1$), where the scalar field rolls down its potential ($\xi_1 > 1$).
    \end{itemize}
    To illustrate these scenarios, we present numerical solutions for $\xi_1 = 0.5$ (phantom case, $\xi = -1$) and $\xi_1 = 1.5$ (quintessence case, $\xi = 1$). Additionally, the `trivial' $\Lambda$CDM case ($\xi_1 = 1$, $\xi=0$) is included for comparison.
\end{enumerate}

The introduction of the parameter $\xi_1$ as an additional degree of freedom—compared to the mirror AdS-to-dS transition model elaborated in previous sections—provides greater flexibility to this class of models, allowing them to explore more features of observational data, particularly BAO and SnIa measurements. This, in turn, enables a test of the generic nature of the simplest $\Lambda_{\rm s}$CDM framework, which features an exact mirror AdS-to-dS transition In the analysis below, we investigate the dynamical properties of the considered models by imposing consistency with both local measurements of $H_0$ (specifically adopting the SH0ES measurement~\cite{Riess:2021jrx}) and features of the Planck-CMB power spectra (see~\cref{sec:A} for details). These impose background-level constraints, given in Eqs.~\eqref{eq:Omega_M0} and \eqref{eq:D_M}, on the pressureless matter density parameter, assuming $h = 0.7304$, which corresponds to the mean value of the SH0ES measurement, see Ref.~\cite{Alestas:2020mvb,Akarsu:2022lhx}. While these constraints do not necessarily ensure full consistency with other datasets, such as BAO and SnIa measurements, they provide reasonable estimates for the model parameters required to satisfy necessary cosmological constraints. This, in turn, allows us to conduct a realistic discussion of the universe’s dynamics within these models. For all considered cases, we first fix the value of $H_0$ and verify that the model satisfies the constraint given in~\cref{eq:D_M}. This calculation automatically determines the redshift $z$ corresponding to the inflection point of the scalar field potential $\widetilde\Omega_{V}$. We then use~\cref{eq:Omega_M0} to compute $\Omega_{\rm{m}0}$ for the given $H_0$. To perform numerical integrations for the selected cases, we set the rapidity parameter to $\nu = 100$ throughout this section.

\begin{figure*}[!t]
\centering
    \begin{tabular}{@{}c@{}}\hspace{-3mm}
	\includegraphics[width=0.75\linewidth]{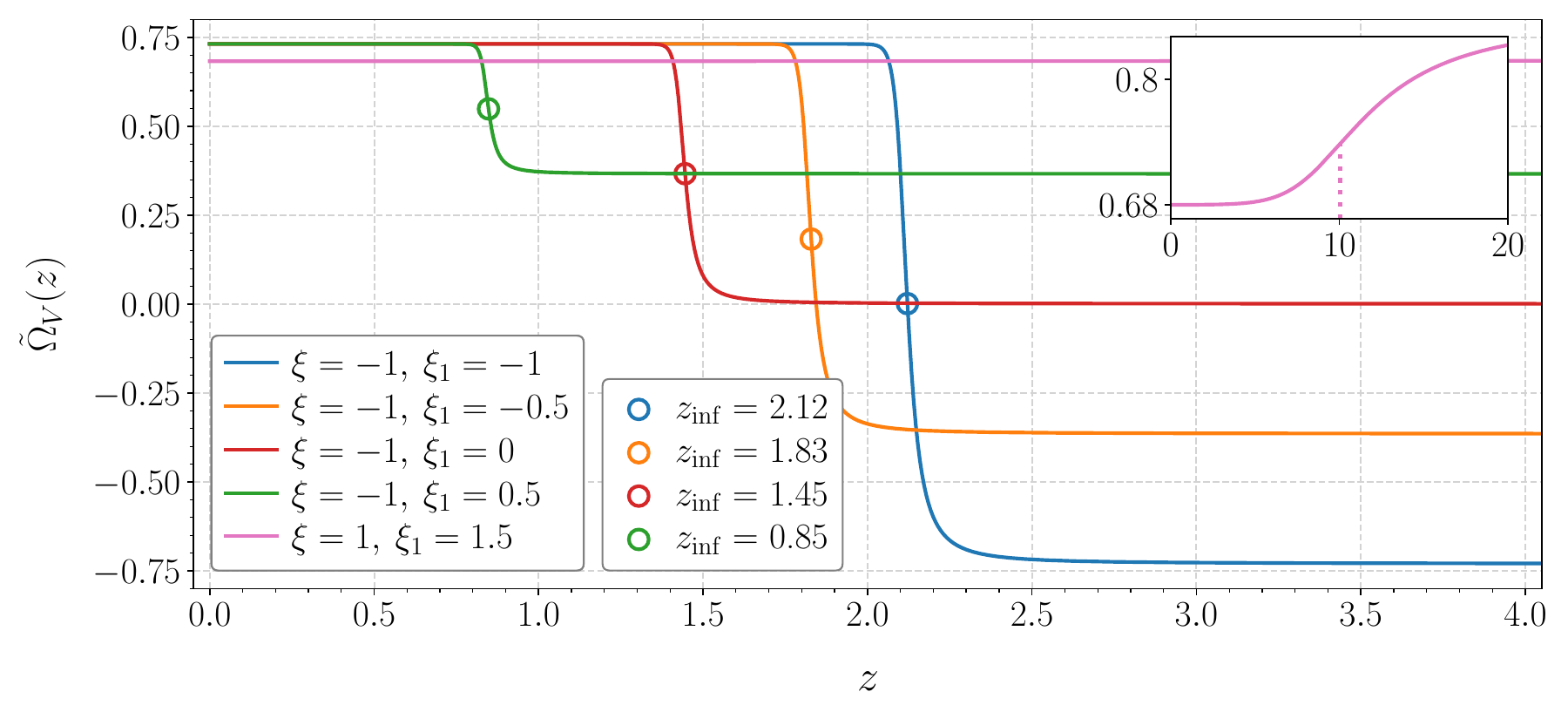}
    \end{tabular}
    \vspace{-3mm}
    \caption{The dimensionless potential energy of the phantom ($\xi=-1$) with parameters $\xi_1=-1,-0.5,0, 0.5$ and quintessence ($\xi=1$) with $\xi_1=1.5$. The marked dots highlight the inflection points of the phantom potentials. Potentials changes sharply at the transition for the phantom. We zoomed a region where quintessence undergoes a smooth transition with the inflaction point at $z\sim 10$. 
    }
\label{fig:sec5_potential} \vspace{7mm}
\end{figure*}

\cref{fig:sec5_potential} illustrates the evolution of the dimensionless potential energy density parameter, $\widetilde\Omega_V$, highlighting the distinct characteristics of the different scalar field cases considered in our model. In the phantom cases ($\xi = -1$), shown for selected values of $\xi_1$ (specifically, $\xi_1 = -1, -0.5, 0,$ and $0.5$), the potentials undergo rapid transitions to the dS phase in the late universe, with the transition epoch lasting for $\Delta z \lesssim 0.2$.   The highlighted circles indicate the redshifts of the inflection points of these potentials, $z_{\rm inf}$, which serve as proxies for the timing of the transition epoch in the universe's expansion history. In the mirror AdS-to-dS case ($\xi_1 = -1$), the inflection point corresponds to the moment when the potential changes sign. Specifically, for this particular case, the inflection point redshift coincides with the transition redshift, i.e., $z_{\rm inf} = z_{\rm t}$. In the $\xi_1 = 0$ case, corresponding to an emerging positive CC, the potential $\widetilde\Omega_V$ prior to the transition is very close to zero (though not exactly zero) over the range of $z\gtrsim z_{\rm inf}$. Our analysis reveals that the deeper the pre-transition potential---relative to its post-transition value (specifically, its present-day value)---the later the transition occurs, meaning the inflection point redshift $z_{\rm inf}$ takes a smaller value. This behavior can be understood as follows. Our model requires a compensatory enhancement in $H(z)$ at low redshifts (i.e., for $z \lesssim z_{\rm inf}$) to match the SH0ES $H_0$ measurement, while ensuring that the comoving angular diameter distance to recombination, $D_M(z_*)$, remains consistent with the value inferred from Planck-$\Lambda$CDM.  Consequently, a shallower pre-transition potential leads to a weaker suppression of $ H(z) $ relative to the Planck-$\Lambda$CDM prediction over the redshift range $ z \gtrsim z_{\rm inf} $, meaning that our model's deviation from Planck-$\Lambda$CDM is reduced compared to cases with a deeper pre-transition potential. To maintain the necessary enhancement in $H(z)$ at low redshifts—ensuring consistency with the SH0ES $H_0$ measurement while preserving the fixed $D_M(z_*)$ integral—the duration of $H(z)$ suppression in the pre-transition era for $\xi_1>-1$ must extend further into the late universe compared to the mirror AdS-to-dS case ($\xi_1 = -1$), meaning that $z_{\rm inf}$ must be smaller than the value $z_{\rm inf} = 2.12$ obtained for the mirror AdS-to-dS case (i.e., $\Lambda_{\rm s}$CDM).  In summary, our model naturally exhibits a strong correlation between the depth of the pre-transition flat potential and the transition redshift: the shallower the potential before the transition, the longer the suppression of $H(z)$ must persist to maintain consistency with the SH0ES $H_0$ measurement, leading to a later transition in the expansion history, i.e., a smaller value of $z_{\rm inf}$.

\begin{figure*}[!t]
\centering
    \begin{tabular}{@{}c@{}}\hspace{-3mm}
	\includegraphics[width=0.75\linewidth]{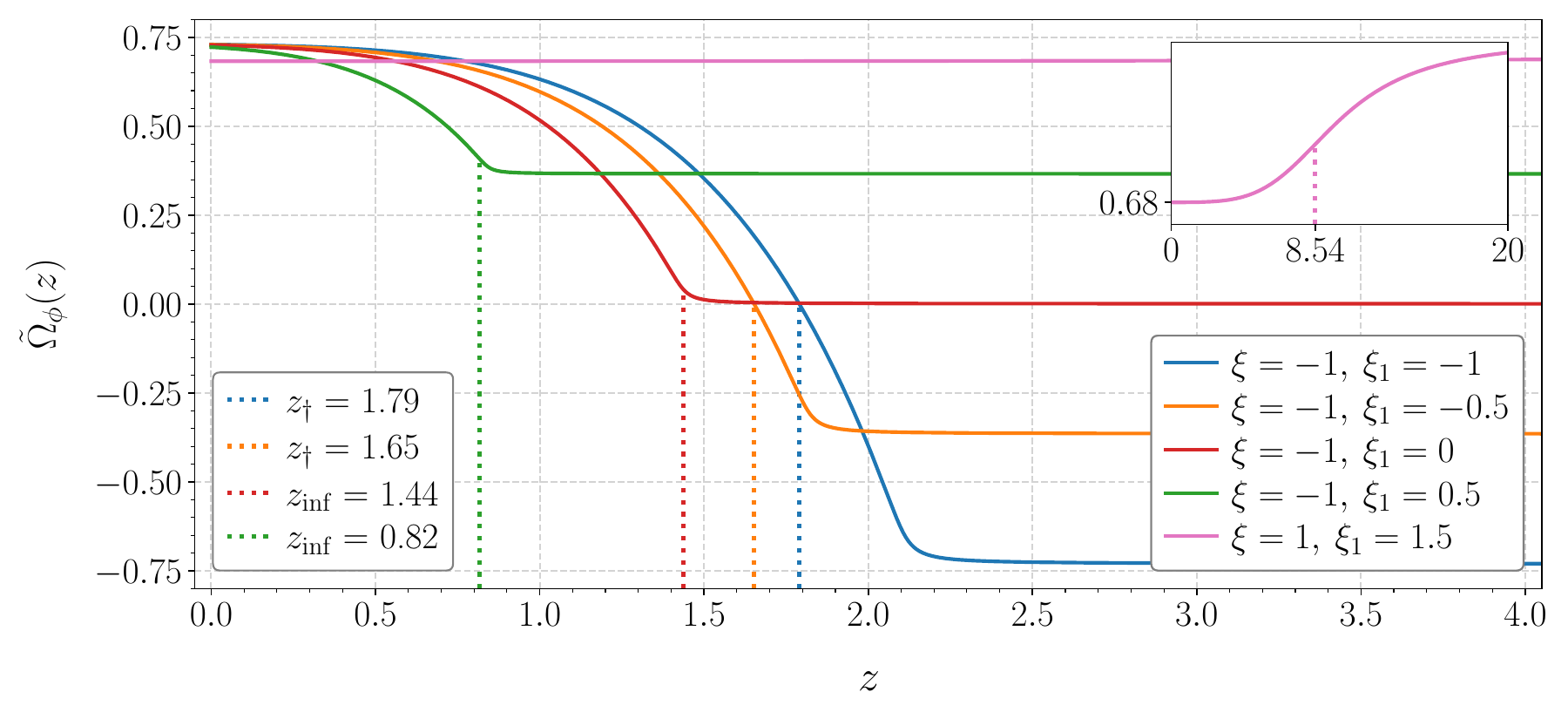}
    \end{tabular}
    \vspace{-3mm}
    \caption{The dimensionless total energy of scalar field. Redshifts $z=0.82, 1.44, 8.54$ correspond to the inflection points of $\widetilde\Omega_\phi(z)$ for green, red and pink curves whereas $z=1.65, 1.79$ indicates times when $\widetilde\Omega_\phi(z)$ crosses zero for orange and blue curves, respectively.}
\label{fig:sec5_total_energy} \vspace{7mm}
\end{figure*}

\begin{figure*}[!t]
\centering
    \begin{tabular}{@{}c@{}}\hspace{-3mm}
	\includegraphics[width=0.75\linewidth]{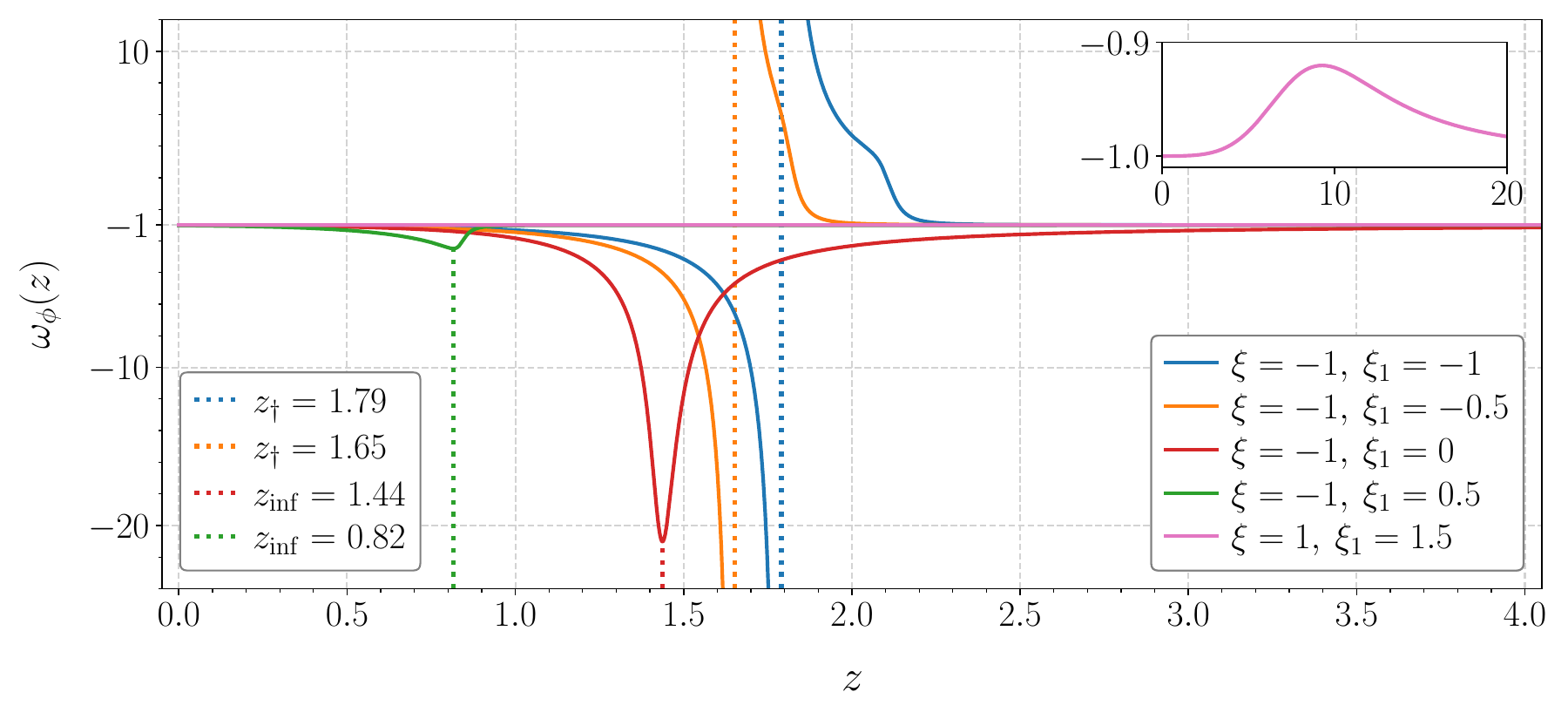}
    \end{tabular}
    \vspace{-3mm}
    \caption{The behavior of the equation of state parameters $\omega_{\phi}$ is defined by the characteristic points of the scalar field total energy $\widetilde\Omega_{\phi}$. In the phantom cases with $\xi_1\geq 0$, the inflection points of $\widetilde\Omega_{\phi}$ correspond to sharp (but smooth) minima of $\omega_{\phi}$. If the total phantom energy crosses zero (that happens for $\xi_1<0$ in the case of blue and orange lines), then the equation of state parameter diverges at this redshift.}
\label{fig:sec5_eq_of_state} \vspace{7mm}
\end{figure*}

\cref{fig:sec5_total_energy} displays the evolution of the scalar field’s dimensionless energy density parameter, $\widetilde\Omega_{\phi}$. We highlight the inflection points of $\widetilde\Omega_{\phi}$ for the cases $\xi_1 = 0$ and $0.5$, while for $\xi_1 = -1$ and $\xi_1 = -0.5$, we mark the points where $\widetilde\Omega_{\phi}$ crosses zero to become positive, i.e., the $z_\dagger$ lines. The blue and orange curves also have inflection points at $z = 2.06$ and $z = 1.77$, respectively, but we do not mark them to avoid cluttering the figure.  We observe that, regardless of its initial value, the phantom scalar field potential rapidly settles into a dS-like flat potential with $\widetilde\Omega_V \approx 0.73$ immediately after the transition epoch (see~\cref{fig:sec5_potential}). In contrast, the evolution of $\widetilde\Omega_{\phi}$ is more gradual---it climbs toward its present-day value less steeply than the potential. This slower evolution is attributable to the kinetic term: as the transition begins, with the steepening of the potential, the scalar field accelerates, and because the phantom field has a negative kinetic term, its kinetic contribution partially counterbalances the increasing potential contribution to the energy density. Once the potential settles into its nearly flat, positive phase, the kinetic term continues to contribute negatively; however, cosmic friction slows down the field, causing the negative kinetic contribution to diminish. As a result, in this phase, $\widetilde\Omega_{\phi}$ gradually approaches the potential value, i.e., $\widetilde\Omega_{\phi} \rightarrow \widetilde\Omega_V \approx 0.73$, as the ratio $-X/V(\phi)=\widetilde\Omega_K/\widetilde\Omega_V$ vanishes with the decreasing redshift. Indeed, for all cases depicted, $\widetilde\Omega_{\phi} \approx \widetilde\Omega_V \approx 0.73$ for $z \sim 0.2$–$0.7$, indicating that our model effectively approximates the standard $\Lambda$CDM behavior in the present-day universe. An important takeaway from our analysis is that while the rapid transition in the potential lasts for $\Delta z \sim 0.2$, the evolution of the energy density extends over a significantly longer period—approximately $\Delta z \approx 0.5$–$1.5$—which influences the behavior of the Hubble function $H(z)$ and the overall expansion history over an extended duration compared to the transition epoch of the potential, which spans only $\Delta z \approx 2$.  Interestingly, a counterintuitive feature emerges: the deeper the pre-transition potential (relative to its post-transition, present-day value), the earlier the energy density "freezes" at $\approx 0.73$, meaning that the model becomes $\Lambda$CDM-like at earlier times. This is understandable since a deeper pre-transition potential necessitates an earlier transition (i.e., a larger $z_{\rm inf}$). However, the key implication is that if late-time data (e.g., for $z \sim 1$–$1.5$) favor $\Lambda$CDM-like behavior, it does not necessarily indicate that $\Lambda$CDM is the preferred model. Instead, it may suggest the presence of significant deviations from $\Lambda$CDM at higher redshifts ($z \gtrsim z_{\rm inf} \sim 1.5$), rather than at low redshifts, as seen in cases with shallower pre-transition potentials in our model.

\begin{figure*}[!t]
\centering
    \begin{tabular}{@{}c@{}}\hspace{-1.5mm}
	\includegraphics[width=0.745\linewidth]{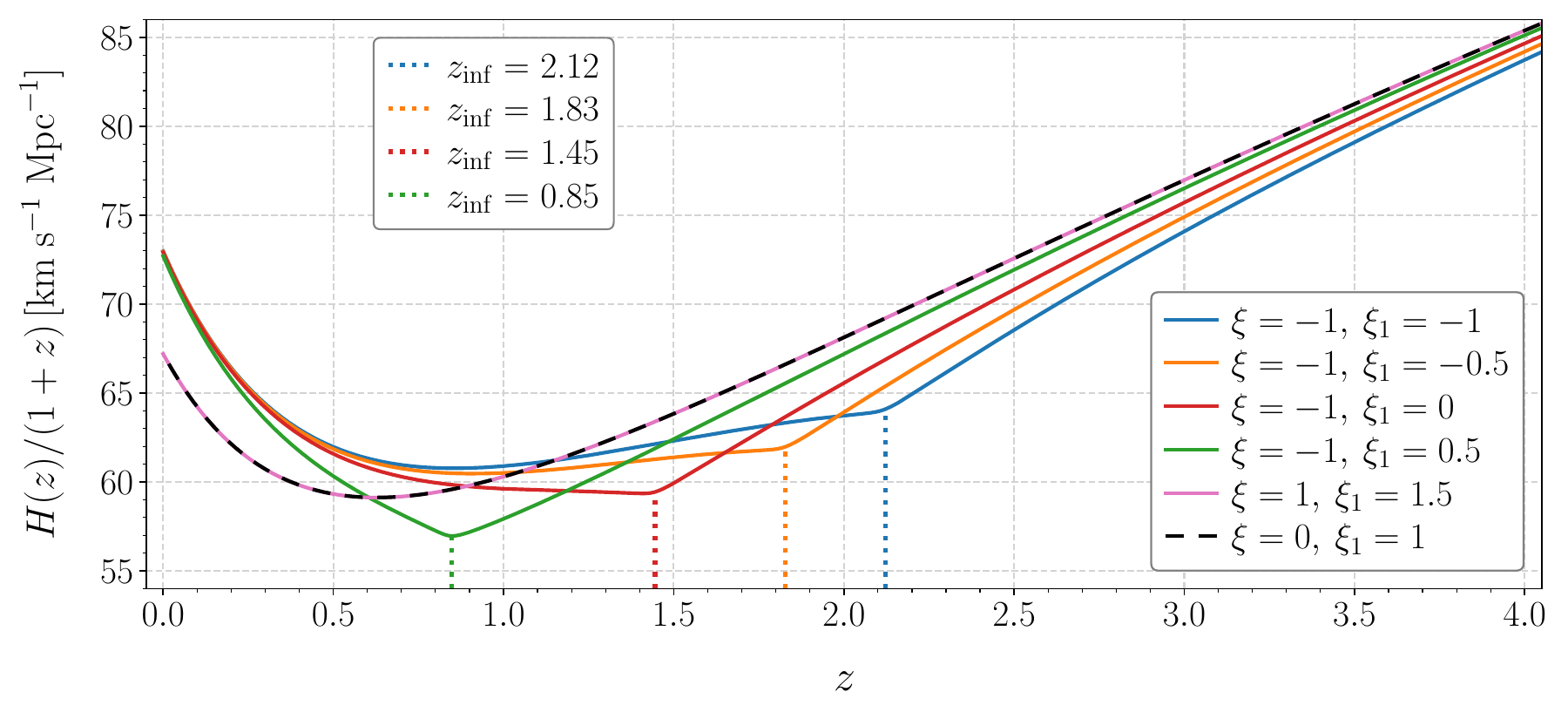}
    \end{tabular}
    \vspace{-3mm}
      \begin{tabular}{@{}c@{}}\hspace{-1.5mm}
	\includegraphics[width=0.75\linewidth]{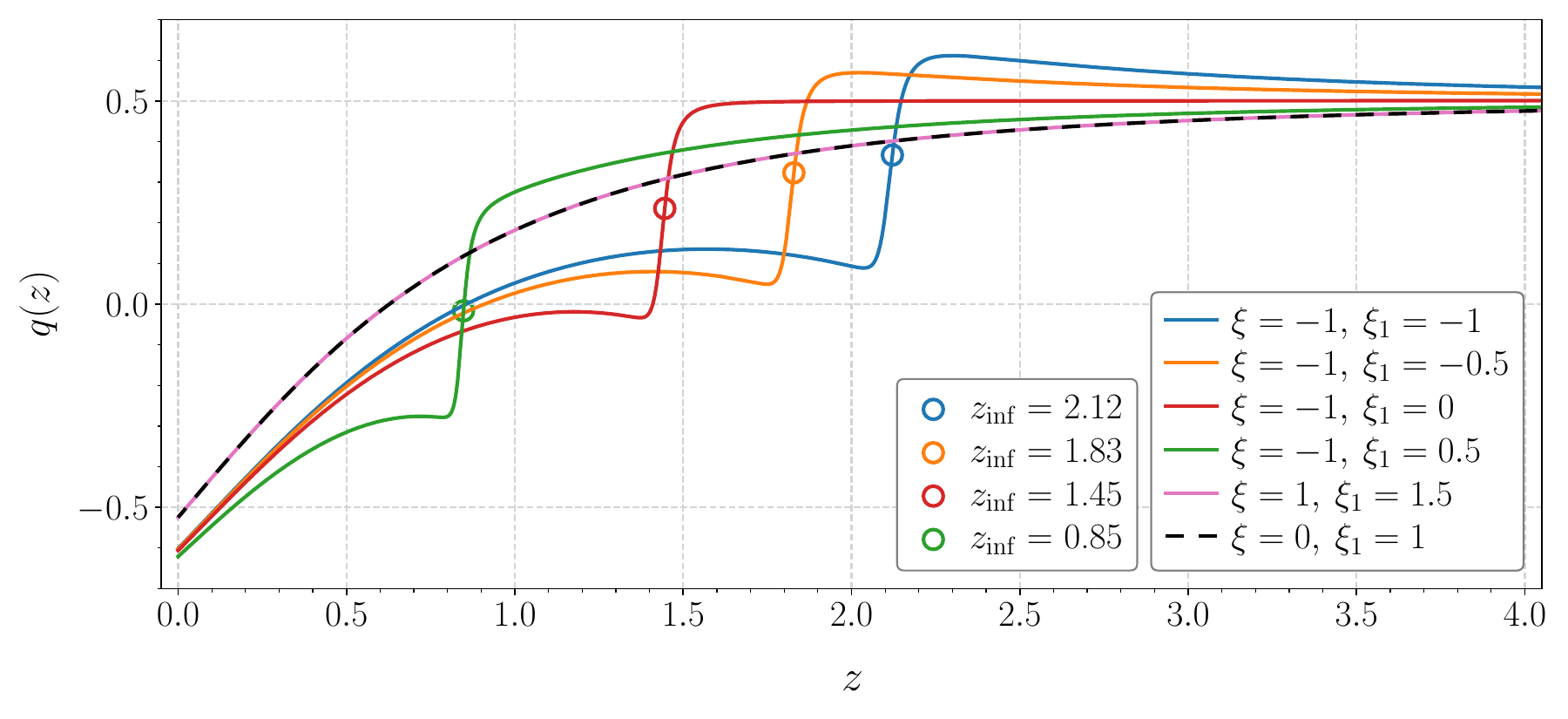}
    \end{tabular}
    \vspace{-3mm}
    \caption{Upper panel: The evolution of the co-moving Hubble parameter, $\dot a=H(z)/(1+z)$, for 
    selected models. All curves are continuous and the inflection points of $\widetilde\Omega_V$ define the characteristic points of $H(z)$. The phantom cases, i.e. $\xi<0$, satisfy the present-day value for the Hubble parameter in accordance with the SH0ES, that is $73.04$. The black dashed line corresponds to $\Lambda$CDM solution which gives $H_0=67.22$ whereas for the quintessence case represented by pink curve $H_0=67.21$ is obtained. Lower panel: The deceleration parameters 
    for the selected models. All curves are continuous and smooth and in the case of the phantom demonstrate the rapid transition. The inflection points of these transitions occur at the same redshifts as the inflection points for the corresponding potentials  $\widetilde\Omega_V$.}
\label{fig:sec5_H_over_zplusone} \vspace{7mm}
\end{figure*}

\cref{fig:sec5_eq_of_state} presents the evolution of the EoS parameter, $\omega_\phi$, for the different cases discussed above. We observe that, sufficiently before and after the transition epoch, $\omega_\phi$ remains arbitrarily close to $-1$, confirming that the our scalar field effectively behaves like a CC in these regimes. However, between these epochs where $\omega_\phi \approx -1$, it undergoes a phase of significant deviation from $-1$, which closely aligns with the transition epoch of $\widetilde\Omega_\phi$ (a broader period compared to the transition epoch of the potential, $\widetilde\Omega_V$). This behavior is expected, as the EoS is determined by both the potential and kinetic terms, similar to the scalar field energy density, viz., $\widetilde\Omega_\phi$. In the cases with $\xi_1 = -1$ and $\xi_1 = -0.5$, where the pre-transition potential is an AdS-like flat potential and the energy density is negative, the EoS parameter exhibits a singularity (pole) when $\widetilde\Omega_\phi$ passes through zero. As discussed earlier for the mirror AdS-to-dS transition $\xi_1 = -1$, this is a “safe” singularity—it occurs solely because $\widetilde\Omega_\phi$ smoothly crosses zero while the pressure remains finite, and the speed of sound remains equal to the speed of light throughout. For these cases, the right branches of the EoS curves exhibit inflection points that coincide with those of $\widetilde\Omega_\phi$, specifically at $z_{\rm inf} = 2.06$ for the blue curve and $z_{\rm inf} = 1.77$ for the orange curve. In contrast, for cases where $\widetilde\Omega_\phi$ does not cross zero (e.g., $\xi_1 = 0$ and $0.5$), no singularity occurs in $\omega_\phi$. Instead, $\omega_\phi$ decreases further into negative values as the transition epoch begins, reaching a minimum at a redshift that coincides with the inflection points of $\widetilde\Omega_\phi$. Then, it starts to increase and eventually becomes arbitrarily close to $-1$ again. Notably, in the emerging CC case ($\xi=-1,\ \xi_1=0$) (see, e.g., Refs.~\cite{Li:2019yem,DeFelice:2020cpt} studying a similar scenario), $\omega_\phi$ tends toward $-1$ at high redshifts, because the initial value of the field derivative is set to zero and the potential, although small, is nonzero. In all cases, for $z \lesssim 0.5$, we find $\omega_\phi \approx -1$, confirming that our model effectively approximates $\Lambda$CDM at late times.

In addition to the phantom cases discussed above, we now consider an example of a quintessence model (\(\xi = +1\)). In a quintessence scenario, the only physically viable option is for the scalar field to roll down its potential. Unlike in the phantom case, where the energy density can transition from negative to positive smoothly, a quintessence field cannot undergo such a transition without encountering a singularity in its energy density (see~\cref{sec:model}). This would violate the FRW background, as \(H(z)^2 \propto \varepsilon(z)\), which is why we exclude this possibility.  On the other hand, it is well known that conventional (i.e., physically viable) quintessence models with \(\varepsilon > 0\) and \(\omega > -1\) tend to exacerbate the \(H_0\) tension. This can be explained as follows: Unlike the standard CC in \(\Lambda\)CDM and phantom models (whether the conventional ones or those in our scenario featuring negative energy densities at high redshifts), the energy density of quintessence increases with redshift. Consequently, it contributes to the growth of \(H(z)\) alongside pressureless matter, leading to an enhancement of \(H(z)\) at high redshifts relative to the Planck-\(\Lambda\)CDM model. To preserve the fixed comoving angular diameter distance \(D_M(z_*)\), this enhancement at high redshifts must be compensated by a suppression of \(H(z)\) at lower redshifts. As a result, in contrast to the phantom case, the quintessence scenario predicts \(H_0\) values that are even lower than those of the Planck-\(\Lambda\)CDM model, thereby worsening the \(H_0\) tension---consistent with the well-known fact that quintessence fields tend to worsen the \(H_0\) tension~\cite{Abdalla:2022yfr}. Returning to the specific quintessence model considered in this work, we examine the case with \(\xi_1 = 1.5\). In this scenario, the transition occurs at significantly higher redshifts compared to the phantom cases—specifically, the inflection point of the potential, \(\widetilde\Omega_V\), is found at \(z_{\rm inf} \approx 10\), while the corresponding inflection point of the energy density, \(\widetilde\Omega_\phi\), is at \(z = 8.54\) (see the magnified views of the pink lines in~\cref{fig:sec5_potential} and~\cref{fig:sec5_total_energy}, respectively). At these high redshifts, the cosmic friction term is substantial, preventing the scalar field from gaining significant speed; consequently, the kinetic term remains subdominant relative to the potential, and the evolution of \(\widetilde\Omega_\phi\) closely follows that of \(\widetilde\Omega_V\). During the transition epoch in the quintessence case, the EoS parameter \(\omega_\phi\) remains above \(-1\) (see~\cref{fig:sec5_eq_of_state}), indicating that the energy density increases with redshift. The magnified view of the pink curve in this figure shows that the maximum value of \(\omega_\phi\) for the considered quintessence model occurs at \(z =9.28\). In our example, the transition occurs at very high redshifts when pressureless matter is still dominant. Consequently, the suppression in \(H(z)\) at low redshifts, for $z\lesssim9$ is minimal (visually negligible), and the worsening of the \(H_0\) value remains minor. For instance, the \(\Lambda\)CDM solution (black dashed line) yields \(H_0 = 67.22\,\mathrm{km\,s^{-1}\,Mpc^{-1}}\), while the quintessence case (pink curve) gives \(H_0 = 67.21\,\mathrm{km\,s^{-1}\,Mpc^{-1}}\). Furthermore, at very low redshifts, the model effectively converges to \(\Lambda\)CDM, as expected.

Lastly, in~\cref{fig:sec5_H_over_zplusone}, we present the evolution of the comoving Hubble parameter, \(\dot{a}(z)\), and the deceleration parameter, \(q(z)\), for all the cases discussed above, providing insight into the expansion kinematics of our models. We focus on the redshift range \(z < 4\), as at higher redshifts all cases effectively converge to the \(\Lambda\)CDM model. For the phantom cases (\(\xi = -1\)), the evolution closely resembles that observed in the mirror AdS-to-dS transition case (\(\xi_1 = -1\)), with the key difference being that the characteristic changes---reflecting the effects of the transition epoch---occur at different redshifts. We observe that for cases with a shallower pre-transition potential, the suppression of \(H(z)\) relative to the \(\Lambda\)CDM model is less pronounced. However, the extended duration of this suppression produces the cumulative effect needed to achieve the compensatory enhancement in \(H(z)\) at low redshifts, thereby pushing \(H_0\) toward the SH0ES measurements~\cite{Riess:2021jrx,Uddin:2023iob,Breuval:2024lsv}. This reinforces the key insight that it is not merely the magnitude of \(H(z)\) suppression before the transition, but also its duration that plays a crucial role in achieving the required \(H_0\). Furthermore, as seen in the deceleration parameter panel, the universe enters its present-day accelerated expansion phase ($q<0$) earlier in all phantom cases than in the \(\Lambda\)CDM model, where this transition occurs at \(z \approx 0.61\). This behavior is expected, as \(H(z)\) in these cases is lower than in \(\Lambda\)CDM before the scalar field's transition epoch (i.e., for \(z \gtrsim z_{\rm inf}\)). However, after the transition (\(z \lesssim z_{\rm inf}\)), \(H(z)\) must eventually increase to reach \(H_0 = 73.04\,\mathrm{km\,s^{-1}\,Mpc^{-1}}\) (the SH0ES mean value~\cite{Riess:2021jrx}) at \(z=0\), rather than the \(H_0 = 67.22\,\mathrm{km\,s^{-1}\,Mpc^{-1}}\) predicted by \(\Lambda\)CDM. Consequently, this requires the universe to begin accelerating earlier (as evidenced by \(q(z)\) crossing below zero earlier than in \(\Lambda\)CDM) and/or to undergo a more rapid acceleration (as indicated by \(q(z)\) remaining lower than in \(\Lambda\)CDM throughout the post-transition era). Specifically, for the cases \(\xi_1 = 0\) (emerging cosmological constant) and \(\xi_1 = 0.5\), the transition occurs so late that the universe enters the accelerated expansion phase right during the transition epoch, at \(z \sim 1.4\) and \(z \sim 0.9\), respectively. The case \(\xi_1 = -0.5\) behaves very similarly to the mirror AdS-to-dS case (\(\xi_1=-1\)) discussed in detail above; both enter the accelerated expansion phase at \(z \sim 0.9\), similar to the \(\xi_1=0.5\) case. However, in contrast to \(\xi_1=0.5\), in both the \(\xi_1=-0.5\) and $-1$ cases, the onset of accelerated expansion ($z\sim0.9$) occurs significantly later than the transition of the phantom field, which takes place at $z_{\rm inf}=1.83$ and 2.12, respectively. In the cases with \(\xi_1 = -1\) and \(\xi_1 = -0.5\), \(q(z)\) rises noticeably above \(1/2\) (the value characteristic of a matter-dominated era) just before the transition. This occurs because the negative energy density of the phantom field begins to compete with dark matter in absolute value, further decelerating the expansion of the universe. In contrast, in the case \(\xi_1 = 0.5\), the parameter \(q\) remains significantly lower than \(1/2\) before the transition, as the positive energy density of the phantom field mitigates the expansion slowdown. Meanwhile, in the case \(\xi_1 = 0\), the phantom energy density remains very close to zero before the transition, resulting in a deceleration parameter \(q\) that remains nearly equal to \(1/2\). In all phantom cases $q(z)$ drops sharply during the transition epoch. However, in all these cases, \(q(z)\) remains higher than in \(\Lambda\)CDM prior to the transition. As a compensatory effect, it subsequently drops below \(\Lambda\)CDM and remains lower throughout the post-transition era. It is worth noting that the condition for \(\dot{H} < 0\) (implying \(q > -1\)) generalizes for arbitrary \(\xi_1\) as follows: \(z_{\rm inf} \gtrsim [4|\xi_1 - 1|(\Omega_{\rm m0}^{-1} - 1)]^{1/3} - 1\). This condition can be easily verified to hold for all considered phantom cases (where we adopt \(\Omega_{\rm m0} \sim 0.27\)), as evidenced by the fact that \(q(z)\) always remains above \(-1\). Finally, for the quintessence case, deviations from \(\Lambda\)CDM are not visually apparent in either \(H(z)\) or \(q(z)\), leading to an \(H_0\) prediction that is nearly identical to that of \(\Lambda\)CDM.

Before concluding, it is important to emphasize that our analysis of the model’s dynamics has been conducted while ensuring consistency with CMB data at the background level. Consequently, all the features observed in these models should be equally compatible with Planck CMB data (at least at the background level), and we do not expect CMB data alone to strongly differentiate between the different cases. However, properly constraining the parameter \(\xi_1\), as well as \(\nu\) (if the analysis is not restricted to fast transitions), would require incorporating additional late-time observational data, such as BAO and SN Ia, to determine whether these models provide a better fit to cosmic history than the \(\Lambda\)CDM scenario. While the present paper focuses on the theoretical formulation and background-level analysis of the model, a comprehensive observational study---including a detailed comparison with late-time datasets and an investigation of perturbations necessary for a full CMB analysis---is left for future work.

\section{Conclusion} 
\label{sec:Conclusion}

In this work, we have developed a robust and physically motivated realization of dynamical dark energy within the $\Lambda_{\rm s}$CDM framework~\cite{Akarsu:2019hmw,Akarsu:2021fol,Akarsu:2022typ,Akarsu:2023mfb,Akarsu:2024eoo,Yadav:2024duq}, offering a promising approach to resolving major cosmological tensions, such as the $H_0$ and $S_8$ tensions. By modeling DE as a minimally coupled scalar field governed by a hyperbolic tangent potential, our construction naturally induces a smooth shift in the effective cosmological constant. This framework accommodates a spectrum of transitions---including from AdS-to-dS, 0-to-dS, to dS-to-dS---with the mirror AdS-to-dS transition serving as a particularly illustrative case that underpins the phenomenology of $\Lambda_{\rm s}$CDM.

A central feature of our model is the employment of a phantom scalar field with a negative kinetic term, which drives a bottom-up transition from an anti–de Sitter (AdS)–like vacuum (negative CC-like DE) at high redshifts to a de Sitter (dS)–like (positive CC-like DE) vacuum at low redshifts. This mechanism is physically well-motivated, as the negative kinetic term reverses the effective force acting on the field, enabling it to climb the potential rather than roll down, as occurs in conventional quintessence scenarios involving a canonical scalar field with a positive kinetic term. Our analysis shows that, despite the inherent challenges associated with phantom fields, the use of a smooth step-like potential---specifically, a hyperbolic tangent potential---successfully avoids typical pathologies such as unbounded energy growth, Big Rip singularities, and violations of the weak energy condition (WEC).

Our numerical integration of the equations of motion—performed while ensuring consistency with the Planck CMB power spectra (fixing the present-day physical matter density \(\Omega_{\rm m0}\, h^2\) and the comoving angular diameter distance to last scattering \(D_M(z_*)\)~\cite{Planck:2018vyg}) and the local SH0ES measurement of \(H_0\)~\cite{Riess:2021jrx,Uddin:2023iob,Breuval:2024lsv}---demonstrates that our model yields a smooth evolution of all key cosmological parameters. In particular, the Hubble parameter \(H(z)\) decreases monotonically as the universe expands (i.e., as  \(z\) decreases), its derivative \(\dot{H}(z)\) remains strictly negative, and the deceleration parameter \(q(z)\) never drops below \(-1\), ensuring that the universe does not enter a super-accelerated expansion phase, even during the transition. Additionally, the phantom field’s energy density and pressure remain finite and evolve smoothly, while the total energy density of the combined phantom and matter system stays positive at all times, thereby preserving the WEC. Moreover, although the phantom field’s equation of state (EoS) exhibits a ``safe'' singularity when its energy density transitions from negative to positive, both the pressure and the speed of sound (which remains equal to the speed of light) behave regularly throughout the transition.

Our detailed analysis of the model leads to several key findings, which we summarize as follows:
\begin{itemize}[nosep,wide]
    \item \textit{Mirror AdS-to-dS Transition:} We have conducted a comprehensive analysis of the mirror (symmetric, i.e., the in magnitude but opposite in sign) AdS-to-dS transition, demonstrating that a phantom field energy density can smoothly evolve from a negative cosmological constant–like regime to a positive one. This provides a solid physical foundation for the \(\Lambda_{\rm s}\)CDM scenario. During the transition, the deceleration parameter \(q(z)\) decreases sharply yet smoothly, reaching a value close to zero. This results in an extended phase of nearly constant expansion lasting approximately from \(z \sim 2\) to \(z \sim 1\). The deceleration parameter \(q(z)\) never crosses -1. Consequently, while the slope of \(H(z)\), i.e., ${\rm d}H(z)/{\rm d}z$, changes rapidly, but it never becomes negative, ensuring that \(H(z)\) remains a monotonically decreasing function—contrasting with the abrupt transition behavior seen in the \(\Lambda_{\rm s}\)CDM scenario.

    \item \textit{Dependence of \(H_0\) on Transition Timing and Rapidity in Mirror AdS-to-dS Transition:}  Our numerical analysis (see~\cref{fig:H0_zdag}) reveals a strong correlation between the transition redshifts \(z_{\rm t}\) (where the potential changes sign) and \(z_\dagger\) (where the phantom field's energy density crosses zero), the rapidity parameter \(\nu\), and the inferred Hubble constant \(H_0\). For selected values of \(\nu\), our model reproduces the SH0ES measurement \(H_0 = 73.04 \pm 1.04~{\rm km\, s^{-1}\, Mpc^{-1}}\)~\cite{Riess:2021jrx} when the transition occurs at \(z_{\rm t} \approx 2.1\) (corresponding to \(z_\dagger \approx 1.8\)). As the transition redshift increases, the predicted \(H_0\) decreases and asymptotically approaches a minimum value \(H_0^{\rm min} = \lim_{z_{\rm t}\to\infty} H_0\) that depends sensitively on \(\nu\). In particular, fast transitions (e.g., \(\nu = 100\) or 1000) yield \(H_0^{\rm min}\) consistent with the Planck-\(\Lambda\)CDM prediction (\(H_0 \approx 67.4 \pm 0.5~{\rm km\, s^{-1}\, Mpc^{-1}}\)), whereas slower transitions (e.g., \(\nu = 10\)–15) yield \(H_0^{\rm min}\) above the Planck value (up to \(68.5~{\rm km\, s^{-1}\, Mpc^{-1}}\)). These findings underscore that both the timing and the rapidity of the DE transition are crucial in setting the present-day expansion rate and provide a viable pathway to alleviate the \(H_0\) tension.
    
    \item \textit{General Transition Scenarios:} In addition to the mirror AdS-to-dS transition, we have systematically explored a spectrum of asymmetric transitions—including AdS-to-dS, 0-to-dS, and dS-to-dS cases—by varying the depth of the pre-transition potential, i.e., \(\xi_1\). Our analysis (with \(\nu=100\) for a fast transition) shows that while the potential sets the overall shift in the effective cosmological constant, the evolution of the DE density is strongly influenced by the kinetic term and cosmic friction. These dynamical effects extend the effective duration of the transition so that the DE density evolves more gradually than the potential itself. Notably, for shallower pre-transition potentials, the transition must occur at lower redshifts (characterized by $z_{\rm inf}$ for general models) to satisfy observational constraints, which prolongs the period during which \(H(z)\) is suppressed and yields a compensatory enhancement at low redshift—resulting in an \(H_0\) value consistent with the local SH0ES measurement. This finding underscores the sensitivity of the late-time expansion history not only to the transition rapidity but also to the depth of the pre-transition potential. Discriminating among these scenarios, as determined by \(\xi_1\), will require late-universe observational data—such as BAO and Type Ia supernovae—in addition to CMB data. 
    
  \item \textit{Observational Consistency and the \(H_0\) Tension:} Our numerical studies reveal a strong correlation between the transition redshift—characterized by \(z_{\rm inf}\) in our general models—and the inferred value of \(H_0\). In our phantom-based scenarios, a relatively late transition (with \(z_{\rm inf}\) ranging roughly from 0.8 to 2, depending on the depth of the pre-transition potential as parameterized by \(\xi_1\) in the range \(-1 \le \xi_1 \le 1\)) leads to an enhanced \(H(z)\) at low redshifts, thereby yielding an \(H_0\) value consistent with the local SH0ES measurement of $73.04\pm1.04~{\rm km\, s^{-1}\, Mpc^{-1}}$~\cite{Riess:2021jrx}. In contrast, analogous quintessence models featuring dS-to-dS transitions predict an \(H_0\) slightly lower than that of standard \(\Lambda\)CDM, and thus they fail to alleviate the \(H_0\) tension.
  
    \item \textit{Safe Singularities and Energy Conditions:} Our analysis further demonstrates that, in the case of phantom fields with AdS-to-dS transitions, the scalar field energy density $\widetilde{\Omega}_{\phi}$ crosses zero at a characteristic redshift, leading to a divergence in the EoS parameter $\omega_\phi$. We show that this divergence is “safe” since it arises solely from the definition of $\omega_\phi$ when $\widetilde{\Omega}_{\phi}$ passes through zero, while the pressure remains finite and the perturbations propagate at the speed of light. Furthermore, the total energy density of the cosmic fluid (comprising both the phantom field and matter) remains positive, and the effective EoS of the total system stays above $-1$, ensuring that the strong form of the WEC is satisfied. In phantom field scenarios featuring 0-to-dS and dS-to-dS transitions, the phantom energy density remains positive at all redshifts, ensuring that the equation-of-state parameter does not exhibit a singularity. Instead, it reaches a pronounced negative minimum at \(z_{\dagger}\) and then asymptotically approaches \(-1\) for redshifts well beyond the transition region.

\item \textit{Dynamical Branches and Transition Feasibility:} We classify the scalar field dynamics into four distinct branches based on the sign of the kinetic term (\(\xi = +1\) for quintessence and \(\xi = -1\) for phantom) and the sign of the energy density. Our analysis reveals that only phantom fields can undergo a smooth transition from a negative-energy (n-phantom) to a positive-energy (p-phantom) regime without encountering unphysical divergences. In contrast, any attempt to achieve a similar transition in the quintessence sector (i.e., from n-quintessence to p-quintessence) results in singularities not only in the EoS parameter---which, even if ``safe'' in the phantom case, can be tolerated---but also, critically, in the energy density itself. Such divergences ultimately disrupt the FLRW geometry, rendering these models physically unrealistic. In other words, within GR, a minimally coupled scalar field cannot smoothly evolve from a state with \(\rho<0\) and \(w<-1\) (n-quintessence) to one with \(\rho>0\) and \(w>-1\); instead, the only viable transition is one in which the field evolves from \(\rho<0\) with \(w>-1\) (n-phantom) to \(\rho>0\) with \(w<-1\) (p-phantom). This fundamental distinction underscores the unique capability of phantom fields to facilitate a viable AdS-to-dS transition in DE.

\end{itemize}

Collectively, these findings establish a robust theoretical foundation for the \(\Lambda_{\rm s}\)CDM framework (and related models). Our work demonstrates that a phantom scalar field endowed with a hyperbolic tangent potential can drive a smooth, controlled (i.e., continuous, with no divergences) AdS-to-dS and dS-to-dS transitions in DE. We call these models as Ph-$\Lambda_{\rm s}$CDM. These models have significant potential to successfully reconcile early-universe CMB constraints with local measurements of \(H_0\), thereby addressing one of the most persistent challenges in modern cosmology.

Looking forward, several important directions remain for further investigation. First, while our study has focused on the background evolution of the model, a full perturbative analysis is required to assess the implications of these transitions for cosmic structure formation and to distinguish among different realizations (e.g., GR-based abrupt \(\Lambda_{\rm s}\)CDM\cite{Akarsu:2021fol, Akarsu:2022typ, Akarsu:2023mfb}, \(\Lambda_{\rm s}\)VCDM~\cite{Akarsu:2024qsi,Akarsu:2024eoo}, \(f(T)\)-\(\Lambda_{\rm s}\)CDM)~\cite{Souza:2024qwd}, and the string-inspired $\Lambda_{\rm s}$CDM$^+$~\cite{Anchordoqui:2023woo,Anchordoqui:2024gfa,Anchordoqui:2024dqc}), e.g., based on their predictions for linear perturbations. Second, a comprehensive confrontation with observational data—particularly from Planck CMB, BAO, Type Ia supernovae, and large-scale structure surveys—will be essential to rigorously test the viability of our model. Finally, extending our framework to incorporate scalar-tensor couplings or other modifications of gravity may provide additional avenues for addressing not only major cosmological tensions such as the \(H_0\) and \(S_8\) discrepancies, but also other emerging observational anomalies~\cite{DiValentino:2021izs,Perivolaropoulos:2021jda,Schoneberg:2021qvd,Shah:2021onj,Abdalla:2022yfr,DiValentino:2022fjm,Akarsu:2024qiq}.

In the present work, we have considered a minimally coupled scalar field with a step-like smooth potential within the framework of GR, and our preliminary observational analysis indicates that our model has strong potential for fitting the data and addressing major cosmological tensions—in line with the demonstrated successes of the \(\Lambda_{\rm s}\)CDM paradigm~\cite{Akarsu:2019hmw,Akarsu:2021fol,Akarsu:2022typ,Akarsu:2023mfb,Akarsu:2024eoo,Yadav:2024duq} in previous studies. Moreover, scalar fields are also integral to broader scalar-tensor theories of gravity, such as Brans–Dicke theory, where the scalar field is non-minimally coupled to gravity and can modulate the effective gravitational coupling (i.e., the Newton constant becomes a function of the scalar field). Consequently, extending our analysis to such frameworks could lead to scenarios in which the step-like smooth transition of the scalar field is accompanied by a corresponding transition in the effective Newton constant ($G_{\rm eff})$. Such a scenario would bridge the \(\Lambda_{\rm s}\)CDM framework~\cite{Akarsu:2019hmw,Akarsu:2021fol,Akarsu:2022typ,Akarsu:2023mfb,Akarsu:2024eoo,Yadav:2024duq} with approaches that propose a ultra-late rapid transition in the gravitational coupling strength, $G_{\rm eff}$, at very low redshifts (e.g., \(z \lesssim 0.1\))~\cite{Alestas:2020zol,Marra:2021fvf,Alestas:2021luu,Perivolaropoulos:2021bds} as a potential solution to the \(H_0\) tension.

In conclusion, our study demonstrates that scalar field Ph-$\Lambda_{\rm s}$CDM models with rapid, smooth transitions in the effective cosmological constant provide a promising and physically motivated framework for resolving major cosmological tensions~\cite{DiValentino:2021izs,Perivolaropoulos:2021jda,Schoneberg:2021qvd,Shah:2021onj,Abdalla:2022yfr,DiValentino:2022fjm,Akarsu:2024qiq}. The \(\Lambda_{\rm s}\)CDM paradigm~\cite{Akarsu:2019hmw,Akarsu:2021fol,Akarsu:2022typ,Akarsu:2023mfb,Akarsu:2024eoo,Yadav:2024duq}, as realized in our work, represents a theoretically minimal yet consistent extension of standard \(\Lambda\)CDM within the framework of GR, offering a compelling path toward a deeper understanding of DE and the late-time evolution of the Universe. Addressing these aspects in future work---through a comprehensive perturbative analysis, further confrontation with observational data from BAO, Type Ia supernovae, and large-scale structure surveys, as well as possible extensions to scalar-tensor theories---will further test the robustness and versatility of our model, ultimately deepening our understanding of DE dynamics and its role in reconciling current cosmological puzzles.

\begin{acknowledgments}
\"{O}.A.\ acknowledges the support of the Turkish Academy of Sciences in the scheme of the Outstanding Young Scientist Award (T\"{U}BA-GEB\.{I}P). This project was also supported by the Hellenic Foundation for Research and Innovation (H.F.R.I.), under the ``First call for H.F.R.I. Research Projects to support Faculty members and Researchers and the procurement of high-cost research equipment Grant" (Project Number: 789). This work was partially supported by the Center for Advanced Systems Understanding (CASUS) which is financed by Germany's Federal Ministry of Education and Research (BMBF) and by the Saxon state government out of the State budget approved by the Saxon State Parliament. A.Z. also expresses gratitude to the CERN Theoretical Physics Department, where part of the work was carried out. This article is based upon work from COST Action CA21136 - Addressing observational tensions in cosmology with systematics and fundamental physics (CosmoVerse), supported by COST (European Cooperation in Science and Technology).
\end{acknowledgments}

\newpage

\appendix
\section{Numerical Method}
\label{sec:A}

To obtain our solutions, we numerically solve Eqs.~\eqref{eq_phi_dless}, \eqref{friedmann_1_dless}, and \eqref{friedmann_2_dless} from the redshift of last scattering, $z = z_* = 1090$, down to $z = 0$. The initial values for the scalar field and its derivative are defined as 
\[
\tilde\phi_{\rm in} \equiv \tilde\phi(z=z_*), \quad \tilde\phi'_{\rm in} \equiv \frac{d\tilde\phi}{dz}(z=z_*),
\]
where a prime denotes differentiation with respect to redshift. The initial Hubble parameter, $\widetilde{h}_{\rm in} \equiv \widetilde{h}(z=z_*)$, is determined from the constraint equation given in Eq.~\eqref{friedmann_1_dless}:
\[
\widetilde{h}_{\rm in}^2 = \frac{\widetilde\Omega_{\rm m}(z_*) + \widetilde\Omega_{\rm r}(z_*) + \widetilde\Omega_V(\tilde\phi_{\rm in})}{1 - \widetilde\Omega_K(z_*,\tilde\phi_{\rm in})},
\]
keeping in mind that $\widetilde\Omega_{\phi}$ depends on $\widetilde{h}^2$ (see Eq.~\eqref{energy_phi_dless}).

We assume that the prerecombination universe is well described by the $\Lambda$CDM model. Thus, using CMB data, we fix the present-day matter density parameter $\Omega_{\rm m0}$ via the constraint (see Table 1 in~\cite{Planck:2018vyg})
\begin{equation}
   \Omega_{\rm m0}\, h^2 = 0.14314, 
\label{eq:Omega_M0}
\end{equation}
where $h \equiv H_0/(100\,{\rm km\,s^{-1}\,Mpc^{-1}})$. For example, taking $H_0 = 73.04\,{\rm km\,s^{-1}\,Mpc^{-1}}$~\cite{Riess:2021jrx} yields $\Omega_{\rm m0} = 0.26831$. In addition, we compute today’s radiation density parameter using
\begin{equation}
    \Omega_{\rm r0}\, h^2 = 2.469 \times 10^{-5} \left[1 + \frac{7}{8}\left(\frac{4}{11}\right)^{\!\!4/3} N_{\rm eff}\right],
    \label{eq:Omega_r0}
\end{equation}
including contributions from both photons and neutrinos (with $N_{\rm eff} = 3.046$ according to the standard model of particle physics).

To determine the value of $\Omega_{\Lambda 0}$ in $\widetilde\Omega_V$ (see Eq.~\eqref{parameters}), we assume that at the present time the potential is very flat (i.e., $\nu(\tilde\phi_0 - \tilde\phi_{\rm c}) \gg 1$) and that the scalar field is nearly frozen (i.e., $(\tilde\phi'_0)^2/6 \ll 1$), so that $\widetilde\Omega_V(z=0) \approx \Omega_{\Lambda 0}$. In this regime, the Friedmann constraint (Eq.~\eqref{friedmann_1_dless}) reduces to
\[
1 \approx \Omega_{\rm m0} + \Omega_{\rm r0} + \Omega_{\Lambda 0},
\]
from which we obtain $\Omega_{\Lambda 0}$ using the values of $\Omega_{\rm m0}$ and $\Omega_{\rm r0}$ determined via~\cref{eq:Omega_M0,eq:Omega_r0}. To validate this approximation, we also define
\[
\mathcal{O} = \Omega_{\rm m0} + \Omega_{\rm r0} + \widetilde\Omega_V(\tilde\phi_0) - 1 - \frac{1}{6}\left(\frac{d\tilde\phi_0}{dz}\right)^2,
\]
and we compute $\mathcal{O}$ at the end of each integration.

We further verify that the comoving angular diameter distance to the last scattering surface is correctly reproduced by
\begin{align}
    D_M(z_*) &= \int_0^{z_*} \frac{c \, dz}{H(z)} = 13869.57 \, \text{Mpc},
    \label{eq:D_M}
\end{align}
where the right-hand side is based on the full CMB data (see the fifth column of Table 2 in~\cite{Planck:2018vyg}) and on the assumption that the $\Lambda$CDM model accurately describes the prerecombination universe~\cite{Akarsu:2022lhx}.

In summary, we choose the values for $\tilde\phi_{\rm in}$, $\tilde\phi'_{\rm in}$, and $\tilde\phi_{\rm c}$ so that $\widetilde{h}_0 = 1$, $\mathcal{O} \approx 0$, and Eq.~\eqref{eq:D_M} is satisfied. The values of $\tilde\phi_0$ and $\tilde\phi'_0$ are then determined by the numerical integration.

Our analysis also shows that the fixation of the cosmological parameters via Eqs.~\eqref{eq:Omega_M0} and \eqref{eq:D_M} implies that, for a given $H_0$, the product $\nu(\tilde\phi_{\rm in} - \tilde\phi_{\rm c})$ is fully determined by the choice of $\nu$. Therefore, the difference $\tilde\phi_{\rm in} - \tilde\phi_{\rm c}$ is fixed for the given $H_0$ and $\nu$, and, without loss of generality, we set $\tilde\phi_{\rm in} = 0$. Consequently, the value of $\tilde\phi_{\rm c} > 0$ is determined automatically. Since the initial value $\tilde\phi_{\rm in}$ is chosen from the region where the potential is nearly flat and cosmic friction is significant (as reflected in the term $3H\dot\phi$ in Eq.~\eqref{fielddyneq}), we can also set the initial velocity $\tilde\phi'_{\rm in} = 0$ without loss of generality.

\newpage

\bibliography{Bibliography}

\end{document}